\documentclass{ptephy}
\usepackage{amsfonts,latexsym,eucal,amsmath,amssymb,times,mathrsfs}
\newcommand{\aap}{Astronomy \& Astrophysics}

\newcommand{\apj}{ApJ}
\newcommand{\apjl}{ApJL}
\newcommand{\apjs}{ApJS}
\newcommand{\apss}{Astrophys. \& Space Science}
\newcommand{\jcap}{JCAP}

\newcommand{\mnras}{MNRAS}
\newcommand{\nar}{New Astron. Rev.}
\newcommand{\nat}{Nature}
\newcommand{\na}{New Astron.}
\newcommand{\physrep}{Phys.Rep.}

\newcommand{\pasj}{PASJ}

\newcommand{\prd}{Phys. Rev. D}

\newcommand{\bm}[1]{\hbox{\boldmath{$#1$}}}

\newcommand{\dd}{{\rm d}}

\newcommand{\uv}[1]{\hat{\mathbf{#1}}}
\newcommand{\fsky}{f_{\rm sky}}
\newcommand{\OL}{\Omega_{\Lambda}}

\newcommand{\mn}{\mu\nu}
\newcommand{\fnl}{f_{\rm NL}}

\begin{document}
%%%-----------------------------------------------------------------%%%
\title{Integrated Sachs Wolfe Effect and Rees Sciama Effect}
\author{\name{\fname{Atsushi} \midname{J.} \surname{Nishizawa}}{1}}
\address{
  \affil{1}{
    Kavli Institute for the Physics and Mathematics of the Universe 
    (Kavli IPMU, WPI), The University of Tokyo, Chiba 277-8582, Japan
  }
  \email{atsushi.nishizawa@ipmu.jp}
}
%%%-----------------------------------------------------------------%%%
\begin{abstract}
  It has been around fifty years since R. K. Sachs and A. M. Wolfe
  predicted 
%an 
the existence of 
anisotropy in the Cosmic Microwave Background (CMB) and
  ten years since the integrated Sachs Wolfe effect (ISW) 
%is 
was
first
  detected observationally. 
%The ISW effect is one of the most interesting physical phenomenon in
%the Universe as it provides 
The ISW effect provides
  us with a unique probe of the accelerating expansion of the Universe.
%  Fortunately wide field galaxy survey of Sloan Digital Sky Survey
%  (SDSS) and Wilkinson Microwave Anisotropy Probe (WMAP) release their
%  data at almost same time in the early 2000's, so that the cross
%  correlation between the large scale structure and CMB has been
%  intensively studied to quest the ISW effect. 
  The cross-correlation between the large-scale structure and CMB has 
  been the most promising way to extract the ISW effect from the data.
  %In this review, we
  %revisit the basic equations around the ISW effect 
  In this article, we review the physics of the ISW effect 
  and summarize recent observational results and interpretations.
\end{abstract}
\subjectindex{Cosmological perturbation theory, Cosmic background
  radiations, Dark energy and dark matter}
%%%-----------------------------------------------------------------%%%
\maketitle
%%%-----------------------------------------------------------------%%%
%%%
%%%
%%%
%%%
%%%-----------------------------------------------------------------%%%
%%%-----------------------------------------------------------------%%%
\section{Overview}
\label{Sec:overview}
%%%-----------------------------------------------------------------%%%
%%%-----------------------------------------------------------------%%%
After the discovery of the isotropic radiation of the Cosmic Microwave
Background (CMB) of the Universe by A. A. Penzias and R. W. Wilson
in 1965 \cite{PenziasWilson:65}, R. K. Sachs and A. M Wolfe
%predicted the 
predict the
existence of 
anisotropy in the CMB associated with the
gravitational redshift in 1967 \cite{SachsWolfe:67}.  They fully
integrate the geodesic equation 
%in the perturbed metric 
in a perturbed Friedman-Robertson-Walker (FRW) metric
in the 
%full
fully
general relativistic framework.  The Sachs-Wolfe (SW) is the first
paper that predicts the presence of the anisotropy in the CMB which 
%is
now plays an important role for 
%the cosmological model constraints
%including dark energy model, modified gravity or non-Gaussianity of the
%primordial fluctuation.
constraining cosmological models, the nature of dark energy, modified
gravity, and non-Gaussianity of the primordial fluctuation.

%let 
Let
us begin 
%with looking back on the citation history brief
by reviewing the history.
%In the first twenty years, the annual citation is around ten, and the
%most of the papers discuss about the theoretical aspects that extend
%the theory to the non-linear collapsed object \cite{ReesSciama:68}, or
%the non standard cosmological model like topological defects
%\cite{KaiserStebbins:84}, or non-Gaussianity but 
In the first twenty years since the Sachs-Wolfe paper, most of 
the works were focused on the extension of the Sachs-Wolfe calculation to
non-linear collapsed object \cite{ReesSciama:68, OstrikerVishniac:86}, 
or non-standard cosmological model such as topological defects
\cite{KaiserStebbins:84}.
%, or non-Gaussian fluctuations {\bf give REFs}.
Partridge and
Wilkinson 1967 first gave a glimpse of the existence of inhomogeneity in
the CMB temperature by using the Dicke Radiometer
\cite{PartridgeWilkinson:67}. They found the temperature excess on the
direction of the known quasar cluster position and considered it as
the Rees-Sciama effect \cite{ReesSciama:68}.  In the age of 
the
COBE
satellite, 
%there was the first pandemic.  
the Sachs-Wolfe paper attracted a huge attention.
Most of the papers 
%are
were
focused on the theoretical prediction that 
%is
was
%related on 
related to
the
observation; prediction of the amplitude of the quadrupole power for
the SW effect. \cite{GoudaSugiyamaSasaki:91, GoudaSugiyama:91,
  GoudaSugiyama:92, BunnSugiyama:95}. 
Crittenden \& Turok 1996 pointed out that the gravitational potential
may decay in the $\Lambda$ dominated Universe at $z < 1$ to produce
the ISW signal \cite{CrittendenTurok:96}. They also proposed a novel
method to detect the ISW effect by cross correlating the large-scale
structure with the CMB. 
Kneissl et al. 1997 made an attempt
to extract the ISW effect by cross correlating the CMB observed by
the COBE with the ROSAT X-ray background \cite{Kneissl+:97}, 
%and 
Boughn \&
Crittenden 
2002 
%use 
used
the NVSS radio galaxies for the cross correlation
\cite{BoughnCrittenden:02}, 
%or 
and
Boughn et al. 1998 used the HEAO1 A2 X-ray
background \cite{BoughnCrittendenTurok:98} but none of them 
%turned to be failed. 
could find the significant detection.
%The second pandemic happened at the WMAP era. 
The Sachs-Wolfe paper has attracted a renewed attention in the WMAP era.
The first detection of the ISW was finally achieved
%with cross correlating with the NVSS, a radio
%galaxy survey and X-ray background measured by HEAO
by cross-correlating the WMAP first-year data with the number count of
radio galaxies from the NVSS data, as well as with the HEAO1 A1 X-ray data
\cite{BoughnCrittenden:04}. Subsequently a lot of detections with
various mass tracers have been reported.  In the early 2000's, 
%much attention has been payed to the cosmological constraints as the ISW
%effect can constrain dark energy model 
much work was focused on obtaining cosmological constraint on dark
energy models from the ISW effect,
while in the late 2000's to present, more and more works 
%study 
studied
%the
various systematic effects which may enter in 
%the 
different ways for different measurement methods.

In this paper, we review the ISW effect from theoretical derivation of
the basic equations to the present cosmological interpretations.
%as a part of {\it CMB cosmology} series.  
The paper is organized as
follows. In section \ref{Sec:theory}, we revisit the derivation of the
CMB anisotropy induced by the perturbation of the background geometry
decomposed into scalar, vector and tensor modes.  In section
\ref{Sec:observation}, we discuss 
%about 
the statistical properties 
of the ISW effect
and the method to measure 
%the ISW effect with 
it in the cross correlation 
%between
with 
the large-scale structure. %and we
We also discuss the possible systematic
effects that 
%mess up 
affect
our interpretations.  In section
\ref{Sec:models}, we provide 
%a 
cosmological applications of the ISW
effect including constraints on dark energy and primordial
non-Gaussianity. %And in 
In section \ref{Sec:summary}, we give a summary.
%and describe the future prospects for the ISW effect from the
%theoretical and observational aspects.

%%%-----------------------------------------------------------------%%%
%%%-----------------------------------------------------------------%%%
\section{Theory of the ISW effect In the Standard Cosmology}
\label{Sec:theory}
%%%-----------------------------------------------------------------%%%
%%%-----------------------------------------------------------------%%%

In this section, we derive the basic equations of the 
%Integrated Sachs Wolfe 
ISW
effect based on the original paper \cite{SachsWolfe:67}.  We
first write the line element in a spatially flat FRW metric,
\begin{equation}
  \label{eq:app3-1.0} 
  ds^2
  =
  g_{\mu\nu} dx^{\mu} dx^{\nu}
  =
  a^2(\tau) \tilde{g}_{\mu\nu} dx^{\mu} dx^{\nu},
\end{equation}
where $a$ is the scale factor that depends solely on the conformal
time $\tau$ and $g_{\mu\nu}$ and $\tilde{g}_{\mu\nu}$ are the metric
and a conformally transformed metric, respectively.
%that have one to one correspondence related with the conformal
%transformation one another. 
The metric consists of perturbed and
unperturbed parts, i.e.
$\tilde{g}_{\mu\nu}=\eta_{\mu\nu}+\tilde{h}_{\mu\nu}$, 
%where
with
the unperturbed metric $\eta_{\mu\nu}={\rm diag}(-, +, +, +)$ and
$\tilde{h}_{\mu\nu}\ll 1$.  
%Here we take all the quantity of order $\mathcal{O}(h^2)$ is
%negligible.  
Here we ignore all the quantity of order $\mathcal{O}(h^2)$ and higher.
Introducing two affine parameters
$\tau$ and $\lambda$ to 
%characterise 
characterize
the photon geodesic in the
$g_{\mu\nu}$ and $\tilde{g}_{\mu\nu}$ metric respectively, we have
$d\tau = a^2 d\lambda$ since the action of the geodesic should be
invariant under the rescaling of $g_{\mu\nu}\rightarrow a^2g_{\mu\nu}$
and $d\tau \rightarrow a^2d\lambda$. To simplify the calculation,
we first work on the $\tilde{g}_{\mu\nu}$ system and then translate 
it to the $g_{\mu\nu}$ system.

The metric perturbation can be decomposed into scalar, vector and
tensor 
%mode
modes.  
%The vector mode perturbation, 
$h_{0i}$ can be divided 
into the contributions from scalar and vector while 
%the tensor mode perturbation, 
$h_{ij}$ can be divided into the 
%contribution 
contributions
from
scalar, vector, and tensor modes as,
\begin{eqnarray}
  \label{eq:app3-1.1}
  \tilde{h}_{00}
  &=& -2A^{\rm (s)} \\
  \label{eq:app3-1.2}
  \tilde{h}_{0i}
  &=&-\partial_i B^{\rm (s)}-B^{\rm (v)}_{i} \\
  \label{eq:app3-1.3}
  \tilde{h}_{ij}
  &=& -2\left[
    D^{\rm (s)}\delta_{ij} - 
    \left(
      \partial_i\partial_j-\frac{1}{3}\nabla^2
    \right)C^{\rm (s)}
  \right] + 
  (\partial_jC^{\rm (v)}_{i} + \partial_iC^{\rm (v)}_{j}) 
  + C^{\rm (t)}_{ij},
\end{eqnarray}
where $A, B, C$ and $D$ are arbitrary functions and superscript with the
parenthesis (s), (v) and (t) stand for the scalar, vector and tensor
quantities respectively. The derivative $\partial_i$ denotes the 3 
dimensional covariant derivative. 
%, i.e.  for arbitrary function $X_i$,
%\begin{eqnarray}
%    X_{i|j} 
%    &\equiv&
%    X_{i,j}+~^{(3)}\Gamma ^{k}_{ij}X_{k},
%    \label{eq:app3-1.4}\\
%    ^{(3)}\Gamma ^{i}_{jk}
%    &\equiv&
%    \frac{1}{2}\gamma^{ia}(\gamma _{ja,k}+\gamma _{ka,j}-\gamma _{jk,a}).
%    \label{eq:app3-1.5}
%\end{eqnarray}
The scalar perturbation is not generated from the vector or tensor
mode and 
%that 
the vector perturbation is not generated from the tensor
mode thus we have constraints as
\begin{eqnarray}
    &&\partial^i B^{({\rm v})}_{i} = 0,
    \label{eq:app3-1.6} \\
    &&\partial^i C^{({\rm v})}_{i} = 0, 
    ~~~ C^{({\rm t})i}_{i} = 0,
    ~~~ \partial^i C^{({\rm t})}_{ij} = 0.
    \label{eq:app3-1.7}
\end{eqnarray}

%%%-----------------------------------------------------------------%%%
\subsection{Scalar mode Linear Perturbation}
\label{sSec:linear}
%%%-----------------------------------------------------------------%%%
We now need to fix a gauge 
degree of 
freedom.  For the scalar perturbation,
the conformal Newtonian gauge
%, or longitudinal gauge 
(longitudinal gauge)
is useful
\cite[e.g.][]{MukhanovFeldmanBrandnberger:92, MaBertschinger:95}.  In
the Newtonian gauge, all the gauge 
degrees of 
freedom 
%is 
are
used to eliminate the
off-diagonal components of the perturbed metric.  Then the variables
are fixed as $B^{\rm (s)}=C^{\rm (s)}=0$, and the metric perturbation
can be fully described by the two scalar quantities of $A=\Phi,
D=\Psi$ which are already gauge invariant.  The metric turns out to be
\begin{equation}
    d{s}^2 
    = 
    a^2[-(1+2\Phi )d\tau +(1-2\Psi )\delta_{ij}dx^idx^j ],
    \label{eq:app3-1.8}
\end{equation}
where $\Phi$ and $\Psi$ are 
%and 
Newtonian potential and curvature 
perturbation respectively. Then we naturally obtain the non 
vanishing connections,
\begin{equation}
  \tilde{\Gamma}_{00}^0 = \Phi', ~~~ 
  \tilde{\Gamma}_{0i}^0=\partial_i\Phi, ~~~ 
  \tilde{\Gamma}_{ij}^0=-\Psi'\delta_{ij},
  \label{eq:app3-1.9}
\end{equation}
where $'\equiv \partial /\partial \tau $. The photon geodesic $x(\lambda)$ 
can be obtained by solving the geodesic equation,
\begin{equation}
  \frac{d \tilde{k}^{\alpha}}{d\lambda}
  =
  \tilde{\Gamma}_{\mn}^{\alpha} \tilde{k}^{\mu} \tilde{k}^{\nu},
  \label{eq:app3-1.10}
\end{equation}
where we introduce the 4-momentum 
$\tilde{k}^{\mu}=\frac{dx^{\mu}}{d\lambda}$. It can be decomposed into 
the unperturbed and perturbed geodesic as 
$\tilde{k}^{\mu}=\bar{k}^{\mu} + \delta \tilde{k}^{\mu}$. The photon energy is 
measured by the observer moving with the fluid, thus the observed energy 
%need 
needs
to be projected with the 4-velocity in the un-tilde $g_{\mn}$ system,
\begin{equation}
  E
  =
  g_{\mn}u^{\mu} k^{\nu}.
  \label{eq:app3-1.11}
\end{equation}
%From the definition of $g_{\mn}u^{\mu}u^{\nu}=1$, and uniform
%component is $u^{\mu}=a^{-1}(1,0,0,0)$, the 4-velocity up to first
%order can be written as $u^{\mu}=a^{-1}(1-\Phi, v^i)$. 
By definition $g_{\mn}u^{\mu}u^{\nu}=1$, the unperturbed component is
$u^{\mu}=a^{-1}(1,0,0,0)$. The 4-velocity is written as $u^{\mu}=a^{-1}(1-\Phi, v^i)$.
The 
%spacial 
spatial
3 dimensional velocity is already first order quantity, hence we do not
need to explicitly solve the spacial part of the geodesic equation,
\eqref{eq:app3-1.10}. The solution for the unperturbed background is
trivial, i.e. $\bar{k}^0=1, \bar{k}^i=e^{i}$, where $\bm{e}=(1, e^{i})$ is
the 4-tangent vector of the geodesic. The time part of the first order
solution of equation \eqref{eq:app3-1.10} is integrated as
\begin{equation}
  \frac{\delta \tilde{k}^0}{a^2}
  = 
  \left. \frac{\delta \tilde{k}^0}{a^2}  \right|_{\tau_*}
  + 2[\Phi(\tau_*)-\Phi(\tau_0)] 
  + \int^{\tau_*}_{\tau_0} (\Phi' + \Psi') \dd\tau,
  \label{eq:app3-1.15}
\end{equation}
where $\tau_*$ denotes the conformal time of the decoupling time, and
$\tau_0$ today.
Now the redshift is defined by the photon energy ratio between emitter
and receiver, $1+z\equiv E(\tau_*)/E(\tau_0)$.%, and
Using the fact that under the rescaling of 
$\tilde{g}_{\mn} \rightarrow g_{\mn}$, 
the
4-momentum scales as 
$\tilde{k}^{\mu} \rightarrow a^{2} k^{\mu}$,
\begin{equation}
    1+z
    =
    \frac{k^{\mu }u_{\mu }|_{\tau_*}}{k^{\nu}u_{\nu }|_{\tau_0}},
    \label{eq:app3.53}
\end{equation}
%where $\tau_*$ denotes the conformal time of the decoupling time, and
%$\tau_0$ today.
 Since the temperature drops with redshift as
$T=T_*/(1+z)$, using equations \eqref{eq:app3-1.11},
\eqref{eq:app3-1.15} and \eqref{eq:app3.53},
observed temperature fluctuation over the sky is
\begin{equation}
  \frac{\delta T}{T}^{\rm (s)}
    = 
    \left.\frac{\delta T}{T}^{\rm (s)} \right|_{\tau_*}
    -\Phi(\tau_0)
    +\Phi(\tau_*) 
    %+ \bm{v}\cdot {\rm {\bf e}}
    + \left[ \bm{v}\cdot {\rm {\bf e}} \right]_{\tau_0}^{\tau_*}
    + \int^{\tau_*}_{\tau_0}
    (\Phi'+\Psi') d\tau,
    \label{eq:app3-1.22}
\end{equation}
apart from the isotropic temperature.  
The first term is the intrinsic photon fluctuation at the last
scattering surface other than those induced by the metric
perturbation; $\delta T/T|_{\tau_*}=\delta_\gamma(\tau_*)/4$.
The second term is the 
%monopole contribution which we can not observe and 
gravitational redshift due to our gravitational potential which is the
monopole contribution and can not be observed.
%the
The
third term represents the
temperature anisotropy caused by the gravitational redshift due to the
potential fluctuations at the decoupling epoch, which is called the
naive or 
%ordinal 
ordinary
Sachs Wolfe effect in the literature. Since it also
has the spacial dependence, we shall write
$\Phi(\tau_*)=\Phi(\tau_*,\uv{n})$ where $\uv{n}$ denotes the angular
position on the sky. 
%In addition to this, we also have 
%$-\delta_\gamma(\tau_*)/4$
%which is an intrinsic photon fluctuation at the decoupling other than
%those induced by the metric perturbation. 
The fourth term is the Doppler 
effect that is induced by the relative motion between the observer and
the CMB last scattering surface. 
%which is seen in the dipole pattern on the sky. 
The final integral term 
represents the temperature anisotropy caused by the time variation of 
gravitational potential integrated along the line of sight, 
%so called the integrated Sachs Wolfe effect. 
and this is the ISW effect.
%In the matter dominated Universe,
%i.e. Einstein de-Sitter Universe, the gravitational potential is
%static in the linear perturbation limit, 
Since the gravitational potential is static in the matter dominated
Universe, i.e. Einstein de-Sitter Universe, the ISW effect vanishes
in the linear perturbation limit. 
Thus the ISW effect induces 
temperature fluctuation at radiation dominated era or dark energy
or curvature dominated Universe. 
The former is called 
%as 
the
early
ISW and the latter 
%as 
the
late ISW effect. We note that from the current
observations the matter radiation equality time, $z_{\rm eq}\simeq 3300$
is well before the decoupling, 
%$z_{\rm eq}\simeq 10^4, z_{\rm dec}=10^3$ 
$z_{\rm dec}=1090$ 
and thus the temperature
fluctuation of 
the
early ISW is 
%projected onto the last scattering surface
%and is not separated from the other sources of the CMB anisotropies.
regarded as a part of the primary anisotropy.
We also note that the careful authors include the visibility function 
to the last scattering surface in the integrand, $e^{-\tau}$, where 
$\tau$ is the optical depth; however, in the flat $\Lambda$CDM Universe,
the redshift where the ISW effect becomes important is at $z<1$, and thus
$e^{-\tau}=1$ is a good approximation.

%%%-----------------------------------------------------------------%%%
\subsection{Vector and Tensor mode Perturbations}
\label{sSec:tensor}
%%%-----------------------------------------------------------------%%%
The photon geodesic is 
%also shifted 
perturbed also
by the vector and tensor 
%mode perturbations. 
modes.
For the vector mode, the geodesic perturbation can be
characterized by $B_i$ and $C_i$. The temperature anisotropy induced
by the vector mode is given by
\begin{equation}
  \frac{\delta T}{T}^{\rm (v)}
  =
  \left. \frac{\delta T}{T}^{\rm (v)}\right|_{\tau_*}
  + \left[ {\mathcal V}^i e_i \right] ^{\tau_*}_{\tau_0} + 
  \frac12 \int ^{\tau_*}_{\tau_0} \! d\tau (\partial_i V_j + \partial_j V_i) e^i e^j,
    \label{eq:vt1}
\end{equation}
where the first term is intrinsic vector type temperature fluctuation,
${\mathcal V}_i$ is 
a
rotational component of the velocity and
$V_i=C'_i + B_i$. In the standard scenario of the inflation, the
vacuum fluctuation generates no 
super-horizon
vector mode perturbation
\cite{LiddleLyth:00}. Even if it is generated with some exotic
mechanisms, vector mode has 
a 
only decaying mode solution 
%so that it can be a negligible contribution to the temperature anisotropy
which can be negligible at later times; thus we shall assume that
the vector mode does not exist \cite{LiddleLyth:93}. 
In practice, it makes negligible contribution to
the CMB temperature observation so we assume it is absent.

%Tensor mode more likely produces the temperature anisotropy.  
The tensor mode, in other words 
the
gravitational wave, is 
%characterized 
given 
by
the traceless transverse rank two tensor, $C_{ij}$ in equation
\eqref{eq:app3-1.3}. The temperature fluctuation induced by the tensor
mode metric perturbation is
\begin{equation}
  \frac{\delta T}{T}^{\rm (t)}
  =
  \left. \frac{\delta T}{T}^{\rm (t)} \right|_{\tau_*}
  -\frac12 \int ^{\tau_*}_{\tau_0} \! d\tau 
  C_{ij}^{\prime {\rm (t)}} e^i e^j.
    \label{eq:vt2}
\end{equation}
The tensor mode decays on the scales smaller than the horizon at the
decoupling, say $\sim 1$ degree, and the significant contribution
comes from the largest scales.  Recently, the BICEP2 experiment
%reported 
report
that they 
%detected 
detect
a signature of large amplitude of the
gravitational wave which is observed by the B-mode polarization power
spectrum of the CMB \cite{BICEP2:14}. 
This is indeed the tensor ISW effect!
The best fit value of the tensor
to scalar ratio is $r=0.16$ after removing the foreground
components.  
%We will discuss how the large amplitude of tensor mode
%perturbation affects the current constraints of the CMB temperature
%fluctuation in the next section.
%As the standard inflation scenario
%predicts that the amplitude of the gravitational wave is small, we
%have not observed the imprint of the existence of the tensor mode. 
%In addition to this, because of the evolutional property of the
%gravitational wave, the tensor mode fluctuation of the large scale
%structure may not generate the significant temperature anisotropies,
%therefore we also neglect the tensor contribution in this paper.

%%%-----------------------------------------------------------------%%%
\subsection{Spectrum of the ISW effect}
\label{sSec:growth}
%%%-----------------------------------------------------------------%%%
\begin{figure}
  \begin{center}
    \includegraphics[width=0.7\linewidth]{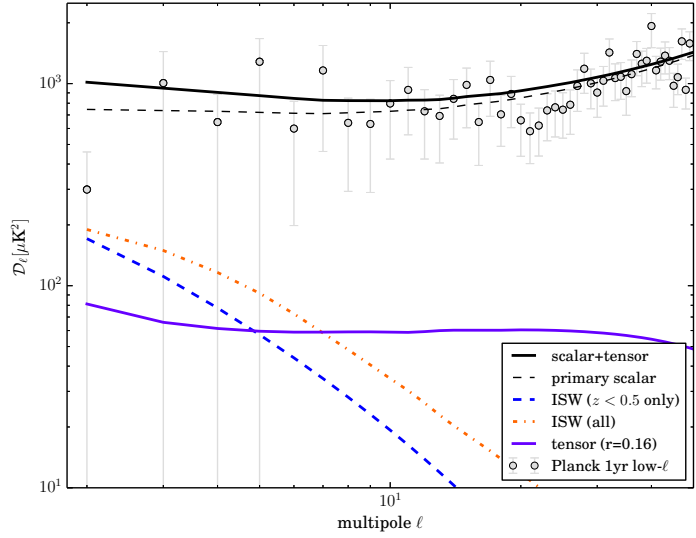}
    \caption{
      Temperature anisotropy at low multipoles.
      The thick solid line shows the total anisotropy integrated
      from today to the last scattering surface including scalar and
      tensor contributions. The thin dashed line shows the scalar
      contribution of the primary anisotropy at the last scattering surface,
      the thick dashed and the dot-dashed lines show contributions
      from the ISW effect generated at $z<0.5$ and $z<100$,
      respectively and the horizontal thick purple line below 
      $100 [\mu {\rm K}^2]$ shows the tensor contribution with $r=0.16$.
      Also shown by points with errorbars is the CMB power spectrum
      obtained by the first year low-$\ell$ Planck data \cite{Planck15:13}.
      \label{fig:isw_spectrum}}
  \end{center}
\end{figure}
%
%As we have seen 
%%at 
%in
%the earlier sections, we only consider the scalar
%mode perturbation. 
In this section, we consider the power spectrum for the scalar mode
perturbation. 
Here we assume that the energy contents of the
Universe have no anisotropic stress, 
%we have $\Phi = \Psi$.
which relates $\Phi$ and $\Psi$ as $\Phi = \Psi$.
Then the integrated Sachs Wolfe effect is written as
\begin{equation}
  \Theta_{\rm ISW}(\uv{n})
  \equiv
  \frac{\delta T_{\rm ISW}(\uv{n})}{T}
  =
  2 \int \frac{\partial \Phi(\uv{n}r, \tau)}{\partial \tau} d\tau.
    \label{eq:spectrum1}
\end{equation}
Now the potential $\Phi$ can be related to the density fluctuation
through the Poisson equation. Working in the Fourier space makes things
simple. In the Fourier space, the Poisson equation in the flat FRW 
Universe is,
\begin{equation}
  \Phi(\mathbf{k},\tau)
  =
  \frac32 \frac{\Omega_m}{a} \left(\frac{H_0}{k}\right)^2
  \left[
    \frac{3aH}{k^2} \theta(\mathbf{k},\tau) + 
    \delta(\mathbf{k},\tau)
  \right]
    \label{eq:spectrum2}
\end{equation}
where $\delta$ and $\theta$ are the density contrast and divergence of
velocity of the matter, respectively and $\Omega_m$ and $H_0$
are the matter density and 
%hubble 
Hubble
parameter today, respectively.  In the Newtonian limit where the scale
of interest is well smaller than the horizon, i.e. $k \gg aH$, equation
\eqref{eq:spectrum2} is reduced to the well known form, 
$\Phi = 3\Omega_m/ 2a (H_0/k)^2\delta$.
We usually expand the
Fourier basis into the spherical harmonics and spherical Bessel function
to 
%yield 
obtain
the full sky expression of the fluctuation 
%to the direction
in a direction
$\uv{n}$, 
\begin{equation}
  \Theta_{\rm ISW}(\uv{n})
  =
  12\pi \Omega_mH_0^2
  \sum_{lm} (-i)^l 
  \int \! d\tau 
  \int \! \frac{dk d\Omega_{\uv{k}}}{(2\pi)^3}
  \frac{\partial }{\partial \tau} \left( \frac{\delta(\mathbf{k},\tau)}{a} \right)
  j_l(kr) Y_{lm}(\uv{n})Y_{lm}^*(\uv{k})
    \label{eq:spectrum3}
\end{equation}
The spherical harmonic counterpart is obtained using the orthogonality
of the spherical harmonics,
\begin{equation}
  \Theta_{lm}
  =
  \int \! \dd\Omega_{\uv{n}} \Theta(\uv{n}) Y_{lm}^*(\uv{n})
  =
  12\pi \Omega_mH_0^2 (-i)^l 
  \int \! d\tau 
  \int \! \frac{dk d\Omega_{\uv{k}}}{(2\pi)^3}
  \frac{\partial }{\partial \tau} \left( \frac{\delta(\mathbf{k},\tau)}{a} \right)
  j_l(kr) Y_{lm}^*(\uv{k}).
    \label{eq:spectrum4}
\end{equation}
The angular power spectrum is then calculated as,
\begin{equation}
  C_l^{\rm ISW}
  =
  \langle \Theta_{lm} \Theta_{lm}^* \rangle
  = 
  \frac{18}{\pi} \Omega_m^2 H_0^4
  \int \! dk P(k) 
  \left[ 
    \int \! dr
    D(f-1){\mathcal H} j_l(kr)
  \right]^2,
    \label{eq:spectrum5}
\end{equation}
where 
%the 
$f$ is the velocity factor, $f\equiv \dd \ln D/\dd \ln a$
and ${\mathcal H}=a'/a$  is the conformal 
%hubble 
Hubble
parameter.
The numerical calculation of equation \eqref{eq:spectrum5} would be
formidable because of the oscillatory behavior of the Bessel function.
The Hankel or also known as Fourier-Bessel transform can be helpful
instead of the direct integration \cite{Hamilton:00, Padmanabhan+:05}.
The Limber's approximation is used in the literature but it is 
%inaccurate up to $\sim 10\%$ 
accurate only up to $\sim 10\%$
on scales larger than $l<10$ 
\cite[e.g.][]{FrancisPeacock:10b}.

Figure \ref{fig:isw_spectrum} shows the power spectra of the primary
CMB temperature fluctuation and the ISW effect for the best fit model
to the Planck first year data combined with the WMAP polarization and
high-$\ell$ CMB experiments \cite{Planck16:13}. We also show the
anisotropy generated from the tensor mode with tensor to scalar ratio
$r=0.16$ which is recently suggested by the discovery of the BICEP2
experiment \cite{BICEP2:14}.  We use the publicly available CAMB code
\footnote{CAMB code is available at http://camb.info/} to compute the
scalar and the tensor spectra. The ISW component can be easily
calculated by the slight modification to the `{\it equations.f90}`.
Most of the signal at $l>20$ comes from
the primary anisotropy while on large scales, significant fraction of
anisotropy is generated at low redshifts, $z<0.5$. We also show the
first year low-$\ell$ Planck data \footnote{Planck data can be
  retrieved from http://www.esa.int/Planck}.

Because the large $r$ value constrained by the B-mode power spectrum,
$C_l^{\rm BB}$ of the BICEP2 enhances the amplitude of the temperature
fluctuation, $C_l^{\rm TT}$ on large scales, it is required to make
some modifications of the model to keep the current observational
constraints from the WMAP or the Planck unchanged: either the smaller
(i.e. negatively larger) value of running of scalar spectral index
$\alpha_s$ \cite{BICEP2:14} or isocurvature component of the initial
fluctuation is required \cite{Kawasaki+:14} to suppress the
large scale power of the CMB temperature power spectrum.  However, the
amplitude of tensor mode
%for
with
$r=0.16$ model is subdominant at $\ell < 10$ compared
to the ISW component.  Thus there still remains some
possibilities that the non-standard gravity models 
or the nonlinearity of the local large-scale structure
can alter the
amplitude of the ISW effect, and thus the constraint on the negatively
large value of the running or the fraction of the isocurvature
perturbation component might be reduced.

%%%-----------------------------------------------------------------%%%
\subsection{Non Linear ISW Effect}
\label{sSec:nonlinear}
%%%-----------------------------------------------------------------%%%
%It has been a mere single year to extend the Sachs-Wolfe formula to
%the collapsed object. 
%Rees and Sciama have been working on the large
%scale clustering of the quasar cluster in 1967 \cite{ReesSciama:67}.
M. J. Rees and D. W. Sciama extended the Sachs-Wolfe calculation to non-linear
collapsed objects in 1968 \cite{ReesSciama:68}. 
In 1967, they were investigating an
apparent large-scale clustering of quasars \cite{ReesSciama:67}
reported by Strittmatter \& Faulkner \cite{StrittmatterFaulkner:66}.
They consider the possibility that 
%the inhomogeneity of the large
%scale structure of the cluster of quasar can make the anisotropy of
%the CMB 
inhomogeneity in the matter distribution inferred from the large-scale
clustering of quasars creates anisotropy in the CMB 
as predicted by Sachs \& Wolfe and estimates the amplitude of
the temperature fluctuation induced by 
%the 
a 
spherically symmetric
collapse of the 
%bounded 
objects in an expanding Universe
\cite{ReesSciama:68}.  
Note that here we use the terminology of {\it
  non-linear} as the 
%non-linear collapsed object 
non-linear density fluctuations
but the 
%geodesic perturbation is still keep first order. 
metric perturbations (hence the geodesic perturbations) are kept at
the first order.
 The second order cosmological
perturbation is treated in 
%the 
a consistent manner in \cite{NohHwang:04,
  Tomita:05a, Tomita:05b,TomitaInoue:08} but here we limit our
discussion to the first order geodesic 
%deviation 
equation and 
%non-linearity is due only to the non-linear structure evolution. 
non-linearity is only included in density perturbations.
Using the Poisson
equation, we see that 
%there has two contributions that have
%gravitational potential temporary changed.
there are two contributions to time evolution of a gravitational potential
\begin{equation}
  \Phi'
  =
  \frac32\Omega_m \left(\frac{H_0}{k}\right)^2 
  \frac{1}{a}
  \left( i\bm{k}\cdot\bm{p} - {\mathcal H}\delta \right).
  \label{eq:rs1}
\end{equation}
Here we have used the continuity equation
$\delta'=i\bm{k}\cdot\bm{p}$, where $\bm{p}$ is the momentum of the
density field $\bm{p}=\bm{v}(1+\delta)$.  From equation
\eqref{eq:rs1}, we see that the ISW and RS 
%effect 
effects
consist of
%two component: 
two components: 
one
proportional to the 
%hubble 
Hubble
flow and 
the other the
momentum of the
object. In the linear perturbation limit, the momentum is 
%merely the velocity
equal to the velocity field
so the linear ISW effect in principle 
%trace 
traces
the statistical
property of the large-scale velocity field \cite{CooraySheth:02}.  In
a weakly non-linear regime or fully non-linear regime, a halo model
approach is used to describe the non-linear time evolution of the
gravitational potential which is originally developed for describing
the non-linear clustering of the dark matter halo
\cite{Seljak:96,MaFry:02,CooraySheth:02}.  The dark matter power
spectrum is described by the sum of 
%the 
two contributions: 
a 
two-halo term
where the pair is 
%selected from the 
in
different halos, and 
a 
one-halo term
where the pair is 
%selected from 
in
the same halo \cite{Seljak:00}.  Once
we provide the mass function
\cite[e.g.][]{ShethTormen:99,Jenkins+:01,Warren+:06,Tinker+:10} and
profile of the dark matter halo \cite{NFW:97} then we can immediately
calculate the non-linear clustering of the dark matter. Similarly, the
velocity field can be decomposed into two components: the velocity due
to the virial motion about the center of mass of its parent halo, and
%those 
that
due to the motion of the parent halo itself
\cite{ShethDiaferio:01} which provides the non-linear momentum power
spectrum \cite{MaFry:02}.

Another approach is the higher order perturbation theory.
As we will see in section \ref{Sec:observation}, the cross correlation
between the ISW and density tracer is useful to isolate the
ISW effect from the CMB \cite{CrittendenTurok:96}.
Then the angular cross correlation power spectrum between the ISW and 
any tracers of the density field can be some function of 
the cross
power spectrum of 
$\Phi'$ and $\Phi$,
\begin{equation}
  P_{\Phi'\Phi}(k)
  =
  \frac94\Omega_m^2 \left(\frac{H_0}{k}\right)^4 
  \frac{1}{a^2}
  \left( P_{\delta\delta'} - {\mathcal H}P_{\delta\delta} \right),
  \label{eq:rs2}
\end{equation}
where we employ the notation, 
$\langle X(\bm{k}) Y^*(\bm{k}') \rangle \equiv (2\pi)^3 
\delta^{\rm D}(\bm{k}-\bm{k}')P_{XY}(k)$.
The continuity equation is written as $\delta' = -\theta(1+\delta)$ 
where $\theta$ is the divergence of the velocity. Then the perturbed
variables $\delta$ and $\theta$ can be expanded in a series,
\begin{align}
  &\delta(\bm{k},\tau)=\sum_n D^n(\tau) \delta_n(\bm{k},\tau), \\
  &\theta(\bm{k},\tau)={\mathcal H}f \sum_n D^n(\tau) \theta_n(\bm{k},\tau),
  \label{eq:rs3}
\end{align}
where 
the
$n$th variable is written in terms of 
%the linear fluctuations as
a product of linear fluctuations as
\begin{align}
  \label{eq:rs4}
  \delta_n(\bm{k},\tau)
  =
  &\int \frac{d^3q_1}{(2\pi)^3}\cdots\frac{d^3q_n}{(2\pi)^3} 
  \delta_1(\bm{q}_1)\cdots\delta_1(\bm{q}_n)
  F_n(\bm{q}_1,\cdots,\bm{q}_n) 
  \delta^{\rm D}\left(\sum_i\bm{q}_j-\bm{k}\right) \\
  \label{eq:rs5}
  \theta_n(\bm{k},\tau)
  =
  - &\int \frac{d^3q_1}{(2\pi)^3}\cdots\frac{d^3q_n}{(2\pi)^3} 
  \delta_1(\bm{q}_1)\cdots\delta_1(\bm{q}_n)
  G_n(\bm{q}_1,\cdots,\bm{q}_n) 
  \delta^{\rm D}\left(\sum_i\bm{q}_j-\bm{k}\right),
\end{align}
where $\delta^{\rm D}$ is the Dirac delta function.
The functions $F_{n}$ and $G_n$ describe the mode coupling between
%the different wavevector which is explicitly given by
different wavevectors, and are explicitly given by
\cite[e.g.][]{Makino+:92,JainBertschinger:94}.
Keeping all the 
%term 
terms
which 
%is 
are 
less than fourth order of either 
$\delta_1$ or $\theta_1$, we have \cite{AJNPhDThesis:08}
\begin{align}
  \label{eq:rs6}
  &P_{\delta\delta}(k)
  =
  D^2 P(k)+2D^4 \int \! \frac{\dd ^3q}{(2\pi)^3}P(q)
  \left[
    P(|\bm{k}-\bm{q}|)F_2^2(\bm{q},\bm{k}-\bm{q})
    +3P(k)F_3(\bm{q}, -\bm{q}, \bm{k})
  \right], \\
  \label{eq:rs7}
  &P_{\delta\delta'}(k)
  =
  {\mathcal H}fD^2 P(k) \nonumber \\
  &\hspace{1.5cm}
  -{\mathcal H}fD^4
  \left[
    P(k)
    \int \! \frac{\dd ^3q}{(2\pi)^3}P(q)
    \left(
      F_3(\bm{q}, -\bm{q}, \bm{k})
      +3G_3(\bm{q}, -\bm{q}, \bm{k})
    \right) \right. \nonumber \\
  &\hspace{2.5cm}
  +2\int \! \frac{\dd ^3q}{(2\pi)^3}
  P(q)P(|\bm{k}-\bm{q}|)
  F_2^2(\bm{q},\bm{k}-\bm{q})G_2^2(\bm{q},\bm{k}-\bm{q}) \nonumber \\
  &\hspace{2.5cm}
  \left.
  +2\int \! \frac{\dd ^3q}{(2\pi)^3}
  \right(
    F_2(\bm{q},\bm{k}-\bm{q})P(q)P(|\bm{k}-\bm{q}|)+
    G_2(\bm{q},\bm{q}-\bm{k})P(k)P(|\bm{k}-\bm{q}|)  \nonumber \\
  &\hspace{2.5cm}
    +F_2(\bm{k},-\bm{q}) P(k)P(q)  \left) \frac{\bm{k}\cdot\bm{q}}{q^2} 
  \right],
\end{align}
where $P(k)$ is 
the 
linear power spectrum of $\delta$. More conveniently,
$P_{\delta\delta'}$ can be approximated by 
$\displaystyle P_{\delta\delta'}(k,\tau)=\frac12 \frac{\partial}{\partial \tau} P_{\delta}(k,\tau)$
\cite{Nishizawa+:08,VerdeSpergel:02}.

\begin{figure}
  \begin{center}
    \includegraphics[width=0.5\linewidth]{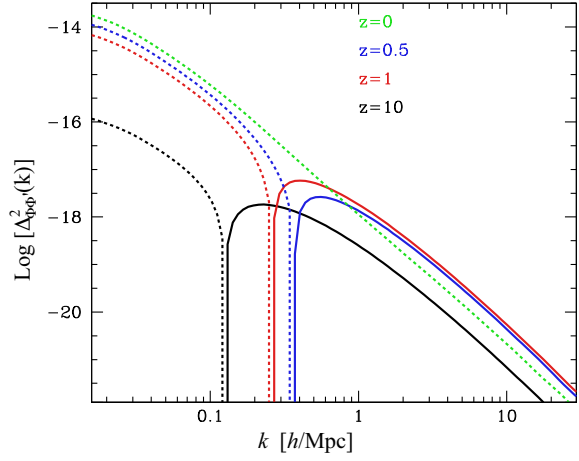}
    \caption{
      The dimensionless cross power spectrum between $\Phi$ and $\Phi'$, 
      $\Delta^2_{\Phi\Phi'}(k)$ at different redshifts computed using
      equations \eqref{eq:rs2}, \eqref{eq:rs6}, and \eqref{eq:rs7}.
      Dashed lines show a negative cross power,
%and solid lines positive.
      while the solid lines positive.
      Figure is 
      %extracted 
      adapted
      from \cite{AJNPhDThesis:08}.
      \label{fig:1loop_phiphidot}
    }
  \end{center}
\end{figure}

Figure \ref{fig:1loop_phiphidot} shows the 3 dimensional cross
correlation power spectrum of $\Phi$ and $\Phi'$ at different
redshifts in $\Lambda$CDM Universe. In the linear regime, 
%at the overdense region gravitational potential is negative $\Phi<0$ 
gravitational potential is negative $\Phi<0$ in the overdense region
and it
decays with time due to the accelerating expansion so that $\Phi'>0$.
Thus $\Phi$ and $\Phi'$ 
%shows 
show an
anti-correlation. %While in 
In the non linear
regime, the gravitational potential grows, i.e.  the potential well
gets deeper thus $\Phi'<0$ which 
%makes $\Phi$ and $\Phi'$ positive correlation.  
gives $\Phi$ and $\Phi'$ a positive correlation. 
At redshift $z=10$, the Universe is close to the matter
dominated so that the linear ISW effect is small and 
the
non-linear RS
effect appears relatively prominently. At 
%more low 
lower
redshifts, the linear
ISW effect is significantly enhanced since the fraction of dark energy 
becomes more substantial but we can still see the transition scale
%at the power turns anticorrelation to correlation at 
%at 
as
the power turns from an anti-correlation to a correlation at 
$k \sim {\mathcal O}(0.1)$ Mpc$/h$. At $z=0$, the linear ISW effect 
dominates 
%over all the scales.
at all scales.

The RS effect generates temperature anisotropy not only through the
contraction and expansion of the structure but also the bulk motion of
the structure perpendicular to the line of sight
\cite{Aghanim+:98,MolnarBirkinshaw:00,Aso+:02,Rubino-Martin+:04}.
Birkinshaw and Gull \cite{BirkinshawGull:83,Birkinshaw:89} point out
that the moving cluster which has transverse velocity to the line of
sight may produce the local dipole structure on the CMB temperature.
In order to explore the complete dynamics, N-body simulation with
ray-tracing method is quite useful. Tuluie \& Laguna
\cite{TuluieLaguna:95, TuluieLagunaAnninos:96} carry out a
ray tracing simulation for the maximum 360 ${\rm Mpc}/h$ box 
%sized
N-body simulation to see the cumulative temperature fluctuation from
$z=100$ to today. They separate the 
%source 
sources of anisotropy into 
%linearISW and non-linear RS, and moving cluster but 
the intrinsic change in gravitational potential of structure and the
transverse bulk velocity of structure; however,
the box size is not
large enough to see the 
%dynamics of large cluster or void that radii
%are more than $100$ Mpc$/h$.  More recently, Cai et
%al.
dynamics of clusters or voids whose sizes are larger than 100 Mpc$/h$.
Cai et al. \cite{Cai+:10}
%revisit the Gigaparsec scale large simulation and construct a full sky
use the Gigaparsec-size simulation to construct a full sky
map of the ISW.  They argue that the non-linear RS 
%or 
and
moving halo
contributions to the total 
%is 
are 
not significant, $<10\%$ but the
%fractional importance 
relative importance (relative to the linear ISW)
is much higher if we go to higher redshift which
is consistent with \cite{Cai+:09, Smith+:09} and with Figure
\ref{fig:1loop_phiphidot}.
This is simply because the linear ISW is negligible at high redshifts
where the Universe is still matter dominated, while the non-linear
RS effect exists regardless of dark energy.

%%%-----------------------------------------------------------------%%%
%%%-----------------------------------------------------------------%%%
\section{Observing the ISW and RS effects}
\label{Sec:observation}
%%%-----------------------------------------------------------------%%%
%%%-----------------------------------------------------------------%%%
The ISW effect is first detected 
%with 
in the CMB measured by the WMAP 
%which is 
cross correlated with the large-scale structure data traced by the
X-ray background radiation and radio galaxies
\cite{BoughnCrittenden:04}.  Several attempts have been made to detect
the ISW effect by cross correlating the CMB map measured by 
the
COBE with
the ROSAT X-ray background \cite{Kneissl+:97}, NVSS radio galaxies
\cite{BoughnCrittenden:02} or HEAO1 X-ray background
\cite{BoughnCrittendenTurok:98} but they 
%turned to be failed. 
could not detect the signal.
However
the non detection of the ISW effect can put 
%a 
an
upper limit on the
amount of dark energy and surprisingly Boughn \& Crittenden
\cite{BoughnCrittenden:02} 
%presents 
present
the limit $\Omega_{\Lambda}<0.74$
which is fairly close to the current limit of dark energy parameter
\cite{Planck17:13}.  Subsequently a number of papers 
%showed up.
appeared.
Table \ref{table:comparison} presents a summary of the detection of
the ISW effect today. Some papers reach consistent 
%conclusion 
conclusions
while
%the other show a remarkable inconsistency each other. 
others show significant inconsistencies.
It is mainly due
to either the wrong statistics used or the contamination of the sample
due to 
%the incomplete subtraction of the possible systematic effects.
an incomplete subtraction of systematics.

In the following sub sections, we 
%will discuss about 
review
the 
%observational side 
observations
of the ISW and RS effects. In sec. \ref{sSec:xcorr}, we define
the cross correlation methods. In sec. \ref{sSec:systematics}, we discuss
%about 
the possible 
%systematics effects that highly affect the
systematic effects that affect the
significance of the detection of the ISW effect.

%%%
\begin{figure*}
  \begin{center}
    \begin{tabular}{cc}
      \includegraphics[width=0.50\linewidth]{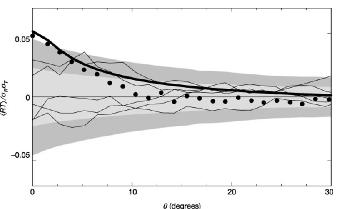}      &
      \includegraphics[width=0.30\linewidth]{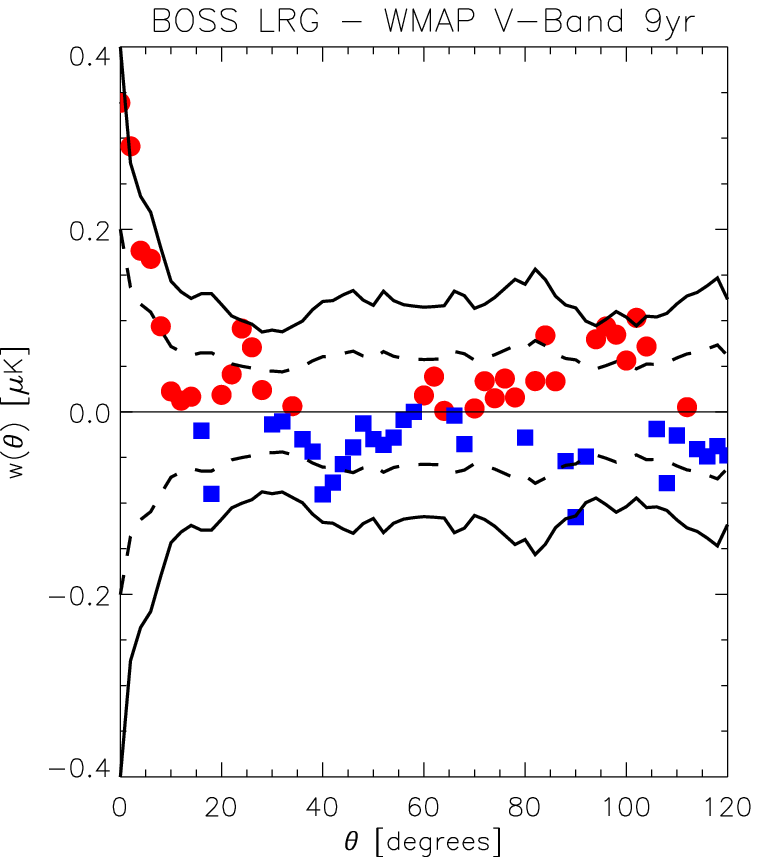} \\
    \end{tabular}
    \caption{
      Left panel shows the first detection of the ISW in 2004 
      by Boughn \& Crittenden \cite{BoughnCrittenden:04}.
      They use the WMAP first-year data cross correlated with
      the NVSS radio galaxies and X-ray data and find the ISW effect
      with 3.0 $\sigma$ detection.
      Right panel shows 
      the latest detection in 2013 by Hernandez-Monteagudo et al. 
      \cite[][but see also
      \cite{Ferraro+:14}]{Hernandez-Monteagudo+:14}. 
       They use the WMAP ninth-year data cross correlated with the
       LRG and CMASS galaxies observed by the BOSS and find the ISW 
       effect with 1.6 $\sigma$ detection.
      %They 
      %%both used 
      %use
      %the WMAP temperature 
      %%map of first and ninth year data release. 
      %maps of first- and ninth-year data releases, 
      %by combining 
      %respectively.
      %%They claims about 3.0 and 1.6 $\sigma$ detection 
      %They claim statistical significance of 3.0 and 1.6 $\sigma$
      %respectively.
      \label{fig:detection_ISW}
    }
  \end{center}
\end{figure*}
%%%

\begin{table}
  \begin{center}
  \scalebox{0.64}{
  \begin{tabular}{lllllll}
    \hline \hline
    Author & CMB & LSS data  & Redshift & Method & Detection & DE Constraints \\
    \hline
    Kneissl et al. '97 \cite{Kneissl+:97} &
       COBE4            &
       ROSAT XRB     &
       N/A                &
       CCF                &
       No detection  &
       --            \\
    Boughn \& Crittenden '98 \cite{BoughnCrittendenTurok:98} &
       COBE4                  &
       HEAO1 XRB          &
       $z \simeq 1$       &
       CCF                      &
       No detection        &
       --                   \\
    Boughn \& Crittenden '02 \cite{BoughnCrittenden:02} &
       COBE4            &
       NVSS              &
       $0 < z < 2$   &
       CCF                &
       No detection  &
       $\OL<0.74$ \\
    Boughn \& Crittenden '04,'05 \cite{BoughnCrittenden:04,BoughnCrittenden:05} &
       WMAP1                 &
       NVSS/HEAO1 XRB &
       $0 < z < 2$          &
       CCF                      &
       2-3$\sigma$       &
       --          \\
    Fosalba \& Gaztanaga '04 \cite{FosalbaGaztanaga:04} &
       WMAP1                   &
       APM                        &
       $z \simeq 0.15$     &
       CCF                         &
       $2.5\sigma$           &
       $\OL=0.8^{+0.06}_{-0.27}(2\sigma)$ \\
    Fosalba et al. '03 \cite{Fosalba+:03} &
       WMAP1              &
       SDSS1(main, LRG)   &
       $z\simeq 0.3, 0.5$ &
       CCF                &
       $2.0, 3.0\sigma$   &
       --                 \\
    &
                     &
       combined      &
                     &
                     &
       $3.6\sigma$   &
       $\OL=0.8^{+0.06}_{-0.11}(2\sigma)$ \\
    Scranton et al. '03 \cite{Scranton+:03} &
       WMAP1         &
       SDSS1(LRG)    &
       $0.3<z_p<0.8$ &
       CCF           &
       $>2\sigma$    &
       --            \\
    Nolta et al. '04 \cite{Nolta+:04} &
       WMAP1         &
       NVSS          &
       $0 < z < 2$   &
       CCF           &
       $2.6\sigma$   &
       $\OL=0.68_{-0.68}(2.2\sigma)$ \\
    Afshordi et al. '04 \cite{Afshordi+:04} &
       WMAP1         &
       2MASS         &
       $0<z<0.2$     &
       APS           &
       $2.5\sigma$   &
       --            \\
    Padmanabhan et al. '05 \cite{Padmanabhan+:05} &
       WMAP1         &
       SDSS4(LRG)    &
       $0.2<z_p<0.6$ &
       APS           &
       $2.5\sigma$   &
       $\OL=0.8^{+0.05}_{-0.19}(2\sigma)$ \\
    Gaztanaga et al. '06 \cite{Gaztanaga+:06} &
       WMAP1         &
       2MASS         &
       $z\sim 0.1$   &
       CCF           &
       $4\sigma$     &
       $\OL=0.70^{\pm 0.05}(1\sigma)^{\rm a)}$ \\
    &
                     &
       APM           &
       $z\sim 0.15$  &
                     &
                     &
       $w=-1.02^{\pm0.17}(1\sigma)^{\rm a)}$ \\
    &
                       &
       SDSS1(main,LRG) &
       $z\sim 0.3,0.5$ &
                       &
                       &
                      \\
    &
                     &
       NVSS+HEAO     &
       $z\sim 0.9$   &
                     &
                     &
                    \\
    Cabre et al. '06 \cite{Cabre+:06} &
       WMAP3            &
       SDSS4(main, LRG) &
       $z\sim 0.3, 0.5$ &
       CCF              &
       $4.4 \sigma$   &
       $\OL\simeq 0.83$ \\
    Giannantonio et al. '06 \cite{Giannantonio+:06} &
       WMAP3         &
       SDSS4(QSO)    &
       $0.1<z_p<2.7$ &
       CCF           &
       $2-2.5\sigma$   &
       $0.075<\Omega_m<0.475(1\sigma)$  \\
    &
       &
       &
       &
       &
       &
       $-1.18<w <-0.76 (1\sigma)$  \\
    Pietrobon et al. '06 \cite{Pietrobon+:06} &
       WMAP3         &
       NVSS          &
       $0<z<2$       &
       WLT           &
       $2.7\sigma$   &
       $0.41<\OL<0.79(2\sigma)$ \\
    Vielva et al. '06 \cite{Vielva+:06} &
       WMAP1         &
       NVSS          &
       $0<z<2$       &
       WLT           &
       $3.3\sigma$   &
       $\OL=0.65^{+0.17}_{-0.20}(1\sigma)$ \\
    &
       &
       &
       &
       &
       &
       $w=-0.70^{+0.35}_{-0.50}(1\sigma)$ \\
    Rassat et al. '07 \cite{Rassat+:07} &
       WMAP3         &
       2MASS XSC     &
       $0<z<0.2$     &
       APS           &
       $<1\sigma$    &
       $\OL=0.85^{+0.04}(2\sigma)$ \\
    McEwen et al. '07 \cite{McEwen+:07} &
       WMAP1         &
       NVSS          &
       $0<z<2$       &
       WLT           &
       $3.9\sigma^{\rm b)}$   &
       $\OL=0.63^{+0.18}_{-0.17}(1\sigma)$ \\
    &
       &
       &
       &
       &
       &
       $w=-0.77^{+0.35}_{-0.36}(1\sigma)$ \\
    Giannantonio et al. '08, '12 \cite{Giannantonio+:08,Giannantonio+:12} &
       WMAP3,7       &
       2MASS         &
       $z\sim 0.2$   &
       CCF           &
       $0.5\sigma,0.7\sigma$   &
        \\
    &
       &
       SDSS6,8(main)&
       $z\simeq 0.3$ &
       &
       $2.2\sigma, 2.2\sigma$   &
       \\
    &
       &
       SDSS6,7(LRG)&
       $z\simeq 0.5$ &
       &
       $2.2\sigma, 2.5\sigma$   &
       \\
    &
       &
       SDSS6(QSO)&
       $0<z<3$&
       &
       $2.5\sigma, 2.3\sigma$   &
       \\
    &
       &
       NVSS&
       $0<z<2$&
       &
       $3.3\sigma, 2.8\sigma$   &
       \\
    &
       &
       HEAO&
       $0<z<2$&
       &
       $2.7\sigma, 2.4\sigma$   &
       \\
    &
       &
       combined&
       &
       &
       $4.5\sigma, 4.4\sigma$   &
       $\Omega_m=0.20^{+0.19}_{-0.11}(2\sigma)$ \\
    Raccanelli et al. '08 \cite{Raccanelli+:08} &
       WMAP3         &
       NVSS          &
       $0<z<2$       &
       CCF           &
       $2.5\sigma$   &
       --            \\
    McEwen et al. '08 \cite{McEwen+:08} &
       WMAP3         &
       NVSS          &
       $0<z<2$       &
       WLT           &
       $\sim 3\sigma$&
       -- \\
    Ho et al. '08 \cite{Ho+:08} &
       WMAP3           &
       2MASS           &
       $0<z<0.2$       &
       CCF             &
       $0.2-1.4\sigma$ &
       \\
    &
       &
       SDSS(LRG) &
       $0.2<z_p<0.6$ &
       &
       $1.3-2.5\sigma$&
       \\
    &
       &
       SDSS(QSO) &
       $0.5<z<2$&
       &
       $0.2-1.4\sigma$&
       \\
    &
       &
       NVSS&
       $0<z<3$&
       &
       $2.9\sigma$&
       \\
    &
       &
       combined&
       &
       &
       $3.7\sigma$&
       $\OL=0.746^{\pm0.09}(1\sigma)$ \\
    Granett et al. '08 \cite{Granett+:08}&
       WMAP5          &
       SDSS6(LRG)     &
       $0.4<z_p<0.75$ &
       STK            &
       $4.4\sigma$    &
       --             \\
    Xia et al. '09 \cite{Xia+:09} &
       WMAP5           &
       SDSS6(QSO)      &
       $0<z_p<3$       &
       CCF             &
       $1.5-2.7\sigma$ &
       $\Omega_m=0.273^{\pm0.019}(1\sigma)^{\rm c)}$ \\
    Hernandez-Monteagudo '10 \cite{Hernandez-Monteagudo:10} &
       WMAP5           &
       NVSS            &
       $0<z<3$         &
       C/A             &
       $\sim 2-3\sigma$&
       --              \\
    Sawangwit et al. '10 \cite{Sawangwit+:10} &
       WMAP5         &
       SDSS5 (LRG)   &
       $0.2<z_s<0.5$ &
       CCF           &
       $0.8\sigma$   &
       --           \\
    &
                     &
       2SLAQ (LRG)   &
       $0.4<z_s<0.7$ &
                     &
       $1.6\sigma$   &
       --            \\
    &
                     &
       AAO (LRG)     &
       $0.5<z_s<0.9$ &
                     &
       $0.4\sigma$   &
       --            \\
    &
                      &
       NVSS           &
       $0<z<2$        &
                      &
       $\sim 2\sigma$ &
       --            \\
    Lopez-Corredoira et al. '10 \cite{Lopez-Corredoira+:10} &
       WMAP5         &
       SDSS7         &
                     &
       CCF           &
       No detection  &
       --           \\
    Francis \& Peacock '10 \cite{FrancisPeacock:10b}&
       WMAP3         &
       2MASS XSC     &
       $0<z_p<0.3$   &
       APS           &
       $\sim 1\sigma$&
       -- \\
%    Illic et al. '11 \cite{Ilic+:11} &
%       WMAP          &
%                     &
%                     &
%       CCF           &
%       $\sigma$   &
%       $\OL=^{+}_{-}(\sigma)$ \\
    Goto et al. '12 \cite{Goto+:12} &
       WMAP7         &
       WISE pre.     &
       $0<z<0.3$     &
       APS           &
       $\sim 3 \sigma$   &
       -- \\
%    Giannantonio et al. '12 \cite{Giannantonio+:12} &
%       WMAP7         &
%                     &
%                     &
%       CCF           &
%       $\sigma$   &
%       $\OL=^{+}_{-}(\sigma)$ \\
    Flender et al. '13 \cite{Flender+:13} &
       WMAP5         &
       SDSS6 (LRG)   &
       $0.4<z_p<0.75$&
       STK           &
       $>3 \sigma$   &
       --            \\
    Planck et al. XIX '13 \cite{PlanckXIX:13}&
       Planck1       &
       BOSS8$^{d)}$  &
       $0<z_s<0.7$   &
       C/A/W         &
       $1.7\sigma$   &
       -- \\
    &
                     &
       SDSS8(main)   &
       $0.1<z_p<0.9$ &
                     &
       $2.0\sigma$   &
       -- \\
    &
                     &
       NVSS          &
       $0<z<2$       &
                     &
       $2.9\sigma$   &
       -- \\
    Ilic et al. '13 \cite{Ilic+:13} &
       WMAP7          &
       SDSS6,7(Void)  &
       $0<z_s<0.7$    &
       STK            &
       $\sim 3\sigma$ &
       --             \\
    Hernandez-Monteagudo et al. '14 \cite{Hernandez-Monteagudo+:14} &
       WMAP9          &
       BOSS8$^{\rm d)}$&
       $0.15<z_p<0.7$ &
       C/A/W          &
       $1.62-1.67\sigma$ &
                         \\
    Kovacs et al. '14 \cite{Kovacs+:14} &
       WMAP7         &
       WISE full     &
       $z\sim 0.15$  &
       APS           &
       $\sim 1.0\sigma$&
       -- \\
    Ferraro et al. '14 \cite{Ferraro+:14} &
       WMAP9          &
       WISE full (gal)&
       $z\sim 0.3$    &
       APS            &
       $\sim 2.6\sigma$&
       -- \\
    &
                      &
       WISE full (AGN)&
       $z\sim 1.1$    &
                      &
       $\sim 1.2\sigma$&
       -- \\
    \hline\hline
  \end{tabular}
  }
  \caption{
    The score sheet for the ISW detection. If the redshift of 
    the 
    large-scale
    structure tracer is measured spectroscopically, we denote the range 
    with $z_s$. 
    %and 
    We use $z_p$ for the range of photometric redshift, whereas 
    %if it is 
    we use $z$ when the redshift is 
    inferred by other methods like integrating the luminosity function,
    %or 
    fitting the amplitude of cross correlation, or partial cross 
    matching with the known redshift sources. %then we denote as $z$.
    The numbers after survey name 
    %stands 
    stand
    for the data release, e.g. 
    WMAP1 is the WMAP first-year data release, and SDSS4 is the SDSS fourth data 
    release and so on.
    \small{
    a) joint constraints with the Type Ia supernovae (SNIa) data.
    b) include a posteriori selection
    c) WMAP5+BAO+SNIa+ISW
    d) CMASS+LOWZ sample
    }
    \label{table:comparison}
  }
  \end{center}
\end{table}

%%%-----------------------------------------------------------------%%%
\subsection{Cross Correlation with LSS}
\label{sSec:xcorr}
%%%-----------------------------------------------------------------%%%
Since the ISW effect is generated 
%where the photon passes
when photons pass
through 
%the
a
time varying gravitational potential of the large-scale structure, 
%it can only be isolated from the other source of the temperature
%fluctuations of the CMB by cross correlating 
it can be detected by cross correlating 
with some tracers of the
large-scale structure. We can write this idea in the equation as
\begin{equation}
  \langle
  \Theta%_{\rm ISW} 
  ~ \delta_{\rm LSS}
  \rangle
  = 
  \langle
  (
  \Theta_{\rm dec}+
  \Theta_{\rm fg}+
  \Theta_{\rm ISW}+
  \Theta_{\rm SZ}+
  \Theta_{\rm lens}+
  \cdots
  ) 
  \delta_{\rm LSS}
  \rangle,
  \label{eq:xcorr-1}
\end{equation}
where $\Theta$ is a temperature fluctuation and $\delta_{\rm LSS}$ is
a
density fluctuation 
%for 
of 
some tracers of 
the 
large-scale structure, e.g.
galaxy number count. 
The 
%source 
sources 
of the CMB temperature fluctuation 
%can be decomposed into pieces where 
include:
$\Theta_{\rm dec}$ is 
%generated at and
the primary anisotropy at or
before the decoupling epoch, 
and 
$\Theta_{\rm fg}$ is any astrophysical
foreground contamination 
%at 
from
solar system or Galactic plane.  The rest
of 
the
terms in equation \eqref{eq:xcorr-1} 
%is 
are
attributed to the cosmological
origin that generates secondary CMB anisotropy.  $\Theta_{\rm ISW}$ is
the 
%integrated Sachs Wolfe 
ISW
effect, %\cite{SachsWolfe:67, ReesSciama:68},
$\Theta_{\rm SZ}$ is 
the
Sunyaev-Zel'dovich (SZ) effect
\cite{SunyaevZeldovich:70} and $\Theta_{\rm lens}$ is 
the
gravitational
lens effect \cite{GoldbergSpergel:99,Hu:00,LewisChallinor:06}.  Here
we assume that the large-scale structure is not correlated with the
primary CMB at 
the
decoupling epoch and we also assume that the ISW
effect and the other secondary CMB sources can 
%well distinguishable
be distinguished
since the ISW effect is only important at largest scales while the
other 
%effect is 
effects are
dominant at much smaller scales. 
%The SZ effect could
%be correlated with ISW effect since they are both attributed to the
%galaxies cluster however the amplitude of the cross correlation is
%smaller than the ISW itself by more than two order of magnitude
%\cite{Cooray:02}.

\subsubsection{Angular Power Spectrum}
The observed quantities used in the literature can be classified into
four: angular power spectrum, angular correlation function, wavelet 
and stacking.
As we observe the CMB temperature fluctuation projected on the sky, it
can be expanded into the spherical harmonic series
\begin{equation}
  a_{lm}^{\rm ISW} = \int \! \dd\Omega_{\uv{n}} ~ \Theta_{\rm ISW}(\uv{n}) Y_{lm}^*(\uv{n}),
  \label{eq:xcorr-2}
\end{equation}
where $\uv{n}$ is a unit vector 
%pointing the two 
pointing toward a two
dimensional position
on the sky, $Y_{lm}$ is the spherical harmonic function, d$\Omega_{\uv{n}}$ is
a volume element of the unit sphere and 
the
integral is over the whole
sky.  

The expanded temperature fluctuation can be cross correlated with the
density tracer X,
\begin{equation}
  a_{lm}^{\rm X}
  =
  \int \! \dd \Omega_{\uv{n}} \delta_{\rm X}(\uv{n}) Y_{lm}^*(\uv{n}),
  \label{eq:xcorr-3}
\end{equation}
where X is the matter tracer field projected on to the sky by
\begin{equation}
  \delta_{\rm X}(\uv{n})
  =
  \int_0^{r_*} \! \dd r \delta_m(\uv{n}r,\tau) W_{\rm X}(r),
  \label{eq:xcorr-4}
\end{equation}
with the projection kernel $W_{\rm X}$. The upper bound of 
the
integral
is defined by the maximum distance of the source that has
%the 
a
non-zero contribution to the projection.
For the galaxy distribution, the projection kernel is,
\begin{equation}
  W_{\rm gal}(r) 
  = 
  b(r) r^2\phi(r) \left[ \int \! \dd r r^2\phi(r) \right]^{-1}
  \label{eq:xcorr-5}
\end{equation}
where $\phi$ is the radial selection function and $b(r)$ is the galaxy
bias which may depend on redshift but not on the scale
(by assumption).  We have
assumed that in the scale where the ISW effect is important, the
galaxy number 
%count 
counts
%is 
are
linearly related to the underlying dark matter
density but strictly speaking 
%the 
%relation breaks in some extent that
%may be a possible source of the systematics. 
the linear relation is valid only on very large scales. It breaks down
on smaller scales, introducing a possible source of systematics.
However 
%in the most of
in most of the ISW studies
%cases, 
%the redshift evolution of the galaxy bias is enough to be taken
%into account \cite{Ferraro+:14}. 
it is typically enough to take into account the redshift evolution of
the galaxy bias only and ignore a possible scale dependence of the
bias. 
The weak lensing convergence is 
%the
another powerful tool to trace the distribution of the dark matter,
which has the kernel \cite[see e.g.][]{BartelmannSchneider:01},
\begin{equation}
  W_{\kappa}(r)
  =
  \frac32 \Omega_m H_0^2
  \frac{r}{a}
  \int_r^{r_*} \! \dd r' p(r') \frac{r'-r}{r'},
  \label{eq:xcorr-6}
\end{equation}
%where $p(r)$ is the galaxy distribution of the shape measured galaxies 
where $p(r)$ is the distribution of galaxies whose shapes are measured
and it is 
normalized as $\int \! \dd r p(r) = 1$.
From the statistical isotropy we can define the angular power spectrum
(hereafter denoted as APS) as
\begin{equation}
  \langle 
  a_{lm}^{\rm ISW} a_{l'm'}^{X*}
  \rangle
  =
  C_{l}^{\rm ISW-X} \delta_{ll'}\delta_{mm'}.
  \label{eq:xcorr-7}
\end{equation}
Here the ensemble average is over 
%the 
possible 
%realization 
realizations
of the
Universe 
%however 
; however, we have only one Universe in reality.  Thus $C_l$ is
estimated by averaging over $2l+1$ modes for each multipole $l$-mode,
\begin{equation}
  \hat{C}_l^{\rm ISW-X}
  =
  \frac{1}{2l+1}
   \sum_{m=-l}^{m=l} a_{lm}^{\rm CMB} a_{lm}^{X*}.
  \label{eq:xcorr-8}
\end{equation}
In the practical observation, 
the
whole sky is not observed.
%Conservatively 
For example,
we mask the region where the signal is not reliable
or significantly contaminated by the foreground such as Galactic 
dust or synchrotron emission or zodiacal light of the solar system.
The extra galactic radio 
%source can 
sources should
also 
%be a contaminant and masked.
be masked.
Then the %full sky $a_{lm}$ is now,
cut sky harmonic coefficient, $\tilde{a}_{lm}$, is given now
\begin{equation}
  \tilde{a}_{lm}
  =
  \int \! \dd\Omega_{\uv{n}} ~ W(\uv{n}) \Theta_{\rm ISW}(\uv{n}) Y_{lm}^*(\uv{n}),
  \label{eq:xcorr-9}
\end{equation}
where $W(\uv{n})$ is 
a 
window function 
( 
or a mask function
)
that takes 1 at
the observed 
%region 
pixel
and 0 at the masked 
%region. 
pixel.
The power spectrum
convolved with the mask function, $\tilde{C}_l\equiv \langle
\tilde{a}_{lm} \tilde{a}_{lm}^{*}\rangle$ is called 
%as 
the Pseudo
power spectrum. It has 
a
smaller amplitude typically by $\tilde{C}_l
\simeq \fsky C_l$ where $\fsky$ is the fraction of sky observed.  The
effect of 
the
mask can be corrected in an unbiased manner with the 
{\it Pseudo $C_l$ estimator} (PCL) \cite{Hivon:02,Efstathiou:04a}, or
{\it quadratic maximum likelihood estimator} (QML)
\cite{Tegmark:97,Efstathiou:04b}.

\subsubsection{Angular Correlation Function}
The cross correlation function (hereafter CCF) is defined as the
Legendre transform of the power spectrum,
\begin{equation}
  C^{\rm ISW-X}(\theta)
  =
  \frac{1}{4\pi}
  \sum_l
  (2l+1)
  C_l^{\rm ISW-X}
  P_l(\cos\theta) 
  b_l^{\rm CMB}
  b_l^{\rm X}
  p_l^2,
  \label{eq:xcorr-10}
\end{equation}
where $p_l$ and $b_l$ are the pixel and beam transfer functions, respectively.
%and conversely,
The inverse transform is
\begin{equation}
  C_l^{\rm ISW-X}
  =
  2\pi \int \! \dd \cos\theta P(\cos\theta) C^{\rm ISW-X}(\theta)
  \left( b_l^{\rm CMB}  b_l^{\rm X}  p_l^2 \right)^{-1}
  \label{eq:xcorr-11}
\end{equation}
The estimator of the correlation function is nothing but
\begin{equation}
  \hat{C}^{\rm ISW-X}(\theta)
  \equiv
  \frac{\sum_{ij} \Theta_{\rm CMB}(\uv{n}_i) \delta_{\rm X}(\uv{n}_j) w_{\rm CMB}(\uv{n}_i) w_{\rm X}(\uv{n}_j)}
  {\sum_{ij} w_{\rm CMB}(\uv{n}_i) w_{\rm X}(\uv{n}_j)},
  \label{eq:xcorr-12}
\end{equation}
where $\cos\theta=\uv{n}_i\cdot\uv{n}_j$. The function $w$ is used
to minimize the variance when the depth of the galaxy survey
or effective sensitivity to the CMB at each pixel is not uniform,
%however it will enhance the cosmic variance 
at the expense of increasing sample variance
\cite{FosalbaGaztanaga:04}.

\subsubsection{Wavelet}
Wavelet analysis is 
%in particular useful at the situations that 
%the structure which has typical scale should be isolated or find
%the signature of the non-Gaussianity of the field. 
useful particularly when signals are localized in both configuration
and frequency space.
For the cross
correlation study the covariance of the wavelet coefficients 
(hereafter WLT) can be used 
\cite{Martinez-Gonzalez+:02,Vielva+:04,Vielva+:06,Pietrobon+:06}.
The WLT covariance 
%at given scale 
at a given scale in configuration space
$R$ is defined as
\begin{equation}
  C_{\Psi}^{\rm ISW-X}(R)
  =
  \frac{1}{N_R} \sum_{\uv{n}}
  \omega_{\rm CMB}(R, \uv{n})  \omega_{\rm X}(R, \uv{n}),
  \label{eq:xcorr-13}
\end{equation}
where $\omega(R,\uv{n})$ is the wavelet coefficient at the position
$\uv{n}$. It can be obtained by 
%convolving the map with the window 
convolving a map with a wavelet
function. Here we assume that the 
%window function is Spherical
%Mexican Hat (SMH),
wavelet function is given by a Spherical Mexican Hat (SMH) wavelet
given by 
\begin{equation}
  \Psi(y,R)
  =
  \frac{1}{\sqrt{2\pi}N(R)}
  \left[ 1+\left(\frac{y}{2} \right)^2\right]^2
  \left[ 2-\left(\frac{y}{R} \right)^2\right]
  e^{-y^2/2R^2},
  \label{eq:xcorr-14}
\end{equation}
where $N(R)$ is a normalization constant, 
$N(R)=R\sqrt{1+R^2/2+R^4/4}$. The distance on the tangent plane is 
given by $y=2\tan(\theta/2)$.
Then the WLT coefficient is,
\begin{equation}
  \omega_X(R,\uv{n})
  =
  \int \! \dd \Omega_{\uv{n}'} X(\uv{n}-\uv{n}')\Psi(\theta',R),
  \label{eq:xcorr-15}
\end{equation}
where the volume element $\dd \Omega_{\uv{n}}$ 
%subtend 
subtends
an infinitesimal solid
angle pointing 
%to 
toward
$\uv{n}=(\theta, \phi)$.
The observed WLT covariance can 
%the 
be
compared with the theoretical
calculation with this formula,
\begin{equation}
  C_{\Psi}^{\rm ISW-X, TH}(R)
  =
  \sum_l \frac{2l+1}{4\pi} p_l^2 \Psi_l^2(R) b_l^{\rm CMB} b_l^{\rm X} C_l^{{\rm ISW-X, TH}},
  \label{eq:xcorr-16}
\end{equation}
where 
%$p_l, b_l$ are the pixel and beam spectrum, 
$\Psi_l$ is the 
spherical harmonic coefficient of the SMH and $C_l^{{\rm ISW}-{\rm X, TH}}$ is the
theoretical prediction of the cross correlation APS between 
the ISW and mass tracer.

The choice of the 
%window 
wavelet
function, equation \eqref{eq:xcorr-14}, is
not unique.  Veilva et al. 2004 and Mukherjee \& Wnag 2004
\cite{Vielva+:04,MukherjeeWang:04} applied SMH 
%window 
wavelet
and Cruz et
al. 2006 \cite{Cruz+:06} used the 
%elliptical SMH 
elliptical Mexican Hat wavelet
to find 
%the cold spot
a prominent cold spot
in the CMB map with more than 2 $\sigma$.  However, Zhang \& Huterer
2010 \cite{ZhangHuterer:10} claimed that 
%the significant cold spot is
%seen only if the SMH window function is used but no significance with
the cold spot is found to be statistically significant only if the SMH
wavelet is used, but not with
the tophat or Gaussian windows.  As we mentioned the WLT is useful for
%find the specific feature of the structure, 
finding localized features in the data
while we should be
%careful for the physical interpretation taking into account the
%degree of freedom that we choose the filter function.
careful for interpreting the results, given the degree of freedom in
choosing the wavelet functions.

\subsubsection{Stacking}
A stacking method is particularly useful 
%to identify the statistical
%photometric or temperature profile 
for measuring the average temperature profile
of the CMB around 
%the cluster or void 
clusters and voids
\cite{Granett+:08,Granett+:09,CaiNeyrinck+:13,PlanckXIX:13,Ilic+:13} 
%and to see the correspondence of 
and for finding the correspondence between 
the temperature fluctuation and the 
specific structure of the density field 
\cite{Granett+:08,Papai+:10,Papai+:11,HernandezMonteagudoSmith:13}.
The stacked CMB temperature can be expressed as
\begin{equation}
  S(R)
  =
  \sum_i^{n} A_i^{-1} 
  \int \! \dd \Omega_i 
  \Delta T(\phi_i, \theta_i) M(\phi_i, \theta_i) \Xi(\theta_i, R)
\end{equation}
where the normalization is $A_i=\int \! \dd \Omega M(\phi_i, \theta_i)$, 
$M$ is the composite mask of the CMB and large-scale structure, and 
$\Xi$ is the filter function. For the compensated filter that 
eliminates the constant offset of the temperature, the filter function
is given as \cite[see e.g.][]{CaiNeyrinck+:13,NishizawaInoue:14},
\begin{equation}
  \Xi(\theta_i, R)
  =
  \left\{
    \begin{array}{rl}
      1, & \theta_i < R \\
     -1, & R \leq \theta_i < \cos^{-1}(2\cos R-1) \\
      0, & {\rm otherwise} \\
    \end{array}
  \right.
  .
\end{equation}
Here we take a local coordinate system 
%that 
in which
the z-axis is pointing to 
the $i$-th cluster or void center with $\theta_i, \phi_i$ and $d\Omega_i$ 
denoting the polar and azimuthal 
%angle 
angles
and the volume element in this local 
coordinate systems,
respectively.
The radius $R$ is the physical scale instead of the angle which 
satisfies the relation $R=\varphi_i \chi(z_i)$ where $\varphi_i$ is the
angle that subtends the 
%size of cluster of void 
size of a cluster or a void
lying at 
a
comoving distance 
$\chi(z_i)$. In other words the CMB temperature profile is rescaled 
before stacking, which 
%makes the significance of the detecting correlation significant 
increases the statistical significance of the correlation signal
\cite{Ilic+:13, NishizawaInoue:14}.
Stacking the CMB at the 
%location of the clusters and voids 
locations of clusters and voids
gives a 
typical profile of the CMB temperature. Since the sign of the ISW 
temperature fluctuation is opposite at the voids and clusters, we 
take the difference 
%of 
between
the stacked temperature at voids and clusters,
i.e.
\begin{equation}
  S(R)
  =
  S_{\rm cluster}(R)-S_{\rm void}(R).
\end{equation}
The interpretation of the significance of the detection of the ISW
with the stacking analysis needs careful treatment. As pointed out in
Ref \cite{HernandezMonteagudoSmith:13}, if we look at the particular
scale, the significance is prominent and have tension to the standard
$\Lambda$CDM prediction but the significance goes down if we combine all
scales; it is so called a posteriori selection effect 
\cite{Bennett.etal:11,Ilic+:13,PlanckXIX:13,HernandezMonteagudoSmith:13}.

%%%-----------------------------------------------------------------%%%
\subsection{systematic errors}
\label{sSec:systematics}
%%%-----------------------------------------------------------------%%%
As seen in table \ref{table:comparison}, 
%the detection significance is
%not consistent each other even if they used the same sample for the
%CMB and large scale structure tracer. 
the reported detection significances vary among papers and are often
not consistent with each other, even if they use the same CMB and
large-scale structure data sets.
It is partly because 
of 
the
difference in the statistics they used and partly because 
of 
the
imperfect control of the systematics. In this section, we discuss
%about the 
possible contamination 
%that 
which
may affect the analysis of the
ISW effect. The main contaminant of the CMB is foreground Galactic
dust and synchrotron emissions. 
On the other hand, the galaxy distribution as a tracer of the
large-scale structure has also uncertainty including the incomplete
star-galaxy separation, magnification bias, 
and
redshift distribution
uncertainties. Furthermore, the SZ effect and point source contamination
can be a source of systematics which is discussed in the later section.

\subsubsection{Magnification bias}
\label{ssSec:mag}
Suppose that we correlate the CMB with the number count of 
%the galaxy.
galaxies.
%The number count of the galaxy reflect the clustering of the galaxy
%while it also generates the apparent clustering. 
The observed number count of galaxies reflects the true, underlying 
clustering of galaxies, while it can also generate an apparent
(artificial) clustering in the sky due to various effects.
The gravitational 
lens effect alters the 
%number count of the galaxy 
number count of galaxies
through two effects.
It locally changes the area of the sky observed hence the number
of galaxies observed.
%while
It also magnifies the light of the 
distant faint galaxies,
%thus 
and thus
enhances the number of galaxies observed
in the vicinity of 
%the massive low-z galaxy 
massive low-z galaxies
\cite[e.g.][]{Broadhurst+:95}.
%Thus the 
Given that we observe the galaxy number density at
a
certain redshift bin, $z_i$, the observed number density can be
written as a sum of two components,
\cite{LoVerde+:07},
\begin{equation}
  \delta_{\rm gal}^{\rm obs}(\uv{n}, z_i)
  =
  \delta_{\rm gal}^{\rm s}(\uv{n}, z_i)+
  \delta_{\rm gal}^{\mu}(\uv{n}, z_i),
  \label{eq:mag-1}
\end{equation}
where $\delta_{\rm gal}^{\rm s}$ is 
the
intrinsic galaxy or, in other words, unlensed galaxy number density and
$\delta_{\rm gal}^{\mu}$ is the magnification bias correction.
Now we make a modification to the equation \eqref{eq:xcorr-5}. 
The underlying three dimensional dark matter fluctuations projected
into 
a
redshift bin $z_i$ with the kernel 
%as
is,
\begin{equation}
  \delta_{\rm gal}^{\rm s}(\uv{n}, z_i)
  =
  \int \! dz b(z) W^{\rm s}(z,z_i) \delta_{\rm m}[\uv{n}\chi(z)],
\end{equation}
where $b(z)$ is a galaxy bias which depends on redshift and $W^{\rm
  s}(z, z_i)$ is a kernel that projects galaxies at $z$ onto
a
redshift bin of $z_i$.
% $z_i$.
The explicit form of the kernel depends on the
model assumed. If we assume that each galaxy has 
a
redshift measured
with 
%the 
a
Gaussian error of $\sigma(z)$, the kernel can be written as
\begin{equation}
  W^{\rm s}(z, z_i)
  =
  \frac12 N(z) 
  \left[
    {\rm erfc} \left( \frac{(i-1)\Delta_z-z}{\sqrt{2}\sigma(z)} \right)
    -
    {\rm erfc} \left( \frac{i\Delta_z-z}{\sqrt{2}\sigma(z)} \right)
  \right],
\end{equation}
where $\Delta_z$ is the width of the redshift bin, erfc is the
complementary error function and $N(z)$ is redshift distribution of
galaxies which satisfies, $N(z) dz = p(r) dr = r^2 \phi(r) dr$
normalized as $\displaystyle \int dz N(z) = 1$.

Now we derive the expression of $\delta^{\mu}_{\rm gal}$.
The gravitational lens deflects light and thus changes the position of
galaxies as
\begin{equation}
  \uv{n}^{\rm s} 
  = 
  \uv{n} + \delta \uv{n},
\end{equation}
where 
the
superscript s denotes the quantity for unlensed galaxies, and
$\delta \uv{n}$ is deflection angle. The magnification can be given by
the Jacobian of the mapping of  $\uv{n}^{\rm s} \rightarrow \uv{n}$: i.e.
\begin{equation}
  A^{-1}
  \equiv 
  \left|
    \frac{\partial \uv{n}^{\rm s}}{\partial \uv{n}}
  \right|.
\end{equation}
The galaxy flux is magnified by
\begin{equation}
  f = A f^{\rm s},
\end{equation}
where $f$ is the observed flux and $f^{\rm s}$ is the intrinsic galaxy flux.
In the weak lensing limit, the magnification can be approximated as $A
\approx 1 + 2 \kappa$, where 
$\kappa$ is the so-called ``convergence field'', which satisfies
$|\kappa| \ll 1$. In this
limit, %the observed galaxies number density 
the observed number density of galaxies
can be expressed in terms
of 
%the unlensed galaxies number density as,
the number density of unlensed galaxies as,
\begin{equation}
  \delta_{\rm gal}^{\rm obs}(\uv{n}, z)
  =
  \delta^{\rm s}_{\rm gal}(\uv{n}+\delta \uv{n}, z)
  +
  (5\alpha(z)-2)\kappa [1 + \delta_{\rm gal}^{\rm s}(\uv{n}+\delta
  \uv{n}, z) ],
\end{equation}
where $\alpha$ is the logarithmic slope of the cumulative number
counts of galaxies at the faint end 
\begin{equation}
  \alpha 
  = 
  \left. 
    \frac{\dd \log N(>m)}{\dd m} 
  \right|_{m_{\rm lim}}.
\end{equation}
This definition of $\alpha$ is only true for galaxy populations that
have a linear slope in the relation between the logarithmic cumulative
number and the magnitude. Though in most cases it is true,  
one can generalize the definition of $\alpha$ by introducing an
efficiency function \cite{Hui+:07}.
In the weak lensing limit, the deflection angle $\delta \uv{n}$ is
also a first order small quantity. Expanding 
$\delta_{\rm gal}^{\rm s}(\uv{n}+\delta \uv{n})$
and keeping the first order term gives
\begin{equation}
  \delta_{\rm gal}^{\rm obs}(z, \uv{n})
  =
  \delta_{\rm gal}^{\rm s}(z, \uv{n})
  +
  (5\alpha-2)\kappa,
\end{equation}
where the second term of the RHS denotes the $\delta^{\mu}_{\rm gal}(z,
\uv{n})$.
Using the expression of the convergence $\kappa$ for the sources 
distributed around the $z_i$ bin, we obtain
\begin{equation}
  \delta_{\rm gal}^{\mu}(\uv{n}, z_i)
  =
  \frac{3\Omega_m H_0^2}{2}(5\alpha(z_i)-2)
  \int \! \frac{dz}{H(z)} g(z, z_i)
  (1+z)\delta_{\rm m}[\uv{n}\chi(z)],
\end{equation}
where $\chi$ is the comoving distance 
and the lensing efficiency function $g$ is given by
\begin{equation}
  g(z, z_i)
  =
  \chi(z)
  \int_z^{\infty} \! dz' \frac{\chi(z')-\chi(z)}{\chi(z')} W(z', z_i). 
\end{equation}
%It is shown that the systematic error on the measurement of the dark
%energy parameter $w$ has not much impact for the low redshift sample
%while it is significant particularly when we cross correlate CMB with 
%the sample at $z>2$ \cite{LoVerde+:07}, thus we have to pay a much
%attention if we conclude the null detection of the ISW with high
%redshift object like QSO as the evidence of the non existence of 
%early dark energy.
It has been shown that the magnification bias in the ISW analysis does
not affect the constraints on dark energy parameters inferred from
low-z samples; however, it can have a significant impact when we
cross-correlate the CMB with galaxy samples at $z>2$
\cite{LoVerde+:07}. Thus, we need to be careful when interpreting the
null detection of the ISW  at high redshift.

\subsubsection{Extinctions}
The Galactic dust extinction may be a major source of 
%the 
systematics
in the measurement of the ISW effect \cite{Giannantonio+:08,
  Giannantonio+:12}.  
%At 
In
the region where the Galactic dust extinction
is high, the 
%galaxies color tends to be 
galaxy colors tend to be
red and alter the large-scale
distribution of the galaxy sample depending on the wavelength
observed. In the 
%highly extinction region, 
high extinction region,
%we also expect radio
%emissions from the Galactic dust thus produce 
%a spurious signal in the cross correlation.
we also expect the observed temperature map to contain
microwave emission from the Galactic dust which then produce a
spurious correlations between 
the
temperature map and the galaxy count
data.
 The advantages to use the near- to mid-infrared 
%is minimizing 
are to minimize
the amount of the dust extinction,
%and it enables us to
which enables us to
mitigate the systematics. 
%As the ISW effect is almost cosmic variance
%limited hence as large area as possible is preferred for the accurate
%analysis, the galaxy distribution or CMB intensity fluctuation should 
%be corrected with correct dust models
Correcting the effect of dust is essential for the cross-correlation
study because it allows us to use more sky area. As the ISW effect
appears only on large angular scales where both the CMB and
galaxy data are almost sample variance limited, increasing the 
usable sky area is the only way to increase statistical significance
of the cross-correlation signal.
\cite{SFD,Yahata+:07,Gold+:11,PlanckXII:13,PlanckXIII:13,Planck_interm_XIV:13}.
However 
%#the model itself 
the dust correction
contains uncertainty and it may affect the
precise measurement. 
%Conservatively many authors masked out the suspicious region of the
%sky. 
A more conservative approach is to simply mask out the contaminated
region of the sky.
%Giannantonio et al. 2013 imposed
%the constraint to exclude the high extinction region of $A_r>0.18$ 
%which remain the available sky region $\fsky=0.22$. They further
%tighten the constraints to $A_r>0.08$ with $\fsky=0.11$ which enlarge
%the statistical error scaled as $\fsky^{-1/2}$. The cross correlation
%signal is systematically increased as the extinction constraint
%is tighten but is still consistent within the 1 $\sigma$ statistical error,
%which means the $A_r>0.18$ threshold is stringent enough compared to
%the statistical accuracy \cite{Giannantonio+:13}.
Giannantonio et al.  \cite{Giannantonio+:14}
exclude the high extinction region with $A_r>0.18$ and $0.08$
which leaves the usable sky fraction of $\fsky=0.22$ and 0.11,
respectively.  The latter has less contamination but larger
statistical error and is statistically consistent with the former to
within 1 $\sigma$. Therefor the former cut appears to be sufficient
for this data set.

\subsubsection{SZ and point sources}
%The 
A 
correlation between the CMB and 
%large scale structure 
the large-scale structure
is also generated
by the SZ effect and point sources. As 
%the 
high energy electron scatters 
%the 
CMB photon through the inverse Compton scattering 
%in the high temperature cluster 
in a high temperature cluster 
of galaxies, 
the
SZ effect distorts the shape
of the blackbody CMB spectrum. For that reason, the temperature change 
induced by 
the
SZ effect depends on the frequency we observe; temperature 
decreases at low frequency while it increases at high frequency and no 
%changes at 217 GHz.
change occurs at 217 GHz.
Thus the SZ signal in the cross correlation 
can in principle be isolated with multi-frequency 
%observation.
observations.
Furthermore, the characteristic 
angular
scale that the SZ effect 
%correlate
correlates
with the large-scale structure is smaller than 
%those 
that
of the ISW effect.
In the harmonics space, the ISW peaks at $l=2$ and 
%vanishes 
becomes negligible
at 
$l=30$ 
%while SZ effect 
while the SZ effect
becomes important at $l>100$ 
\cite{Scranton+:03,Afshordi+:04,HillSpergel:13}.
%The point source 
The point sources
found in the CMB map 
%should be a galaxy. 
are extra-galactic sources such as AGNs and dusty galaxies.
They are
bright in 
%radio or microwave 
low (AGNs) and high (dusty galaxies) frequency bands.
%and also in visible or infrared
%where the many galaxy surveys use. 
They are also visible in optical and infrared bands, where
many galaxy survey data are available.
%It may cause a false signal of the cross correlation however the
%scales the point source becomes prominent is also small compared to
%the ISW 
They also produce a correlation between the CMB and large-scale
structure; however, it can be distinguished from the ISW signal
because point sources produce a correlation at much smaller scales
than that of the ISW effect
\cite{Afshordi+:04, Huffenberger+:06, Eriksen+:07}.
%However if we use the correlation function in the real space,
%we should be careful that correlation function mixes different
%physical scales thus point source and SZ effect does affect
%the ISW signal for the cross correlation measurement 
This discrimination by scale is possible only for the power spectrum in
harmonics space but impossible for the correlation function in the
configuration space because, in the harmonic space, multipole scales
where ISW and point sources signals become prominent are different
while in the configuration space, different angular scales and the
different physical origins do not have one to one correspondance
\cite{Hernandez-Monteagudo:10,Hernandez-Monteagudo+:14}.
Although the point sources can be removed by masking if they are
resolved by CMB experiments, 
%there still remains unresolved galaxies 
%lights integrated along the line of sight which may brings the 
%extra-correlation signal in the cross correlation.
unresolved galaxies may still bring an extra-correlation signal.

%%%-----------------------------------------------------------------%%%
%%%-----------------------------------------------------------------%%%
\section{Application to the Cosmology}
\label{Sec:models}
%%%-----------------------------------------------------------------%%%
%%%-----------------------------------------------------------------%%%

%%%-----------------------------------------------------------------%%%
\subsection{Constraints On Dark Energy in $\Lambda$- and w-CDM model}
\label{sSec:de}
%%%-----------------------------------------------------------------%%%
\begin{figure}
  \begin{center}
    \includegraphics[width=0.8\linewidth]{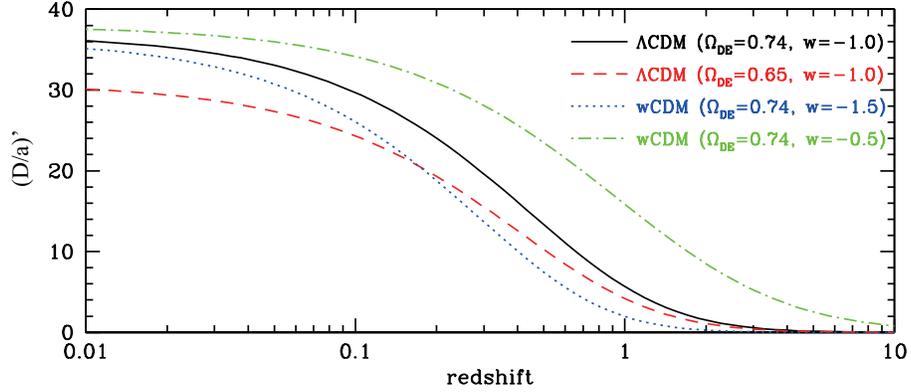}
    \caption{
      Dependence of the ISW effect on the cosmological models,
      $(D/a)'$, 
      as a function of redshift is shown. The less amount of
      dark energy makes Universe more close to the EdS, and thus 
      produces weaker ISW effect (black solid and red dashed). Less 
      negative equation of state, $w$ (green dot-dashed) makes dark 
      energy dominate at earlier epoch and enhances the ISW effect.
      That model can be tested by the cross correlation with high-z
      objects like QSOs or radio galaxies.
      \label{fig:isw_cosmology}
    }
  \end{center}
\end{figure}

As first pointed out by Crittenden \& Turok 1996
\cite{CrittendenTurok:96}, the ISW effect can be used 
%for the unique
as a powerful
probe of dark energy. In this section we describe the background
Universe by the flat FRW metric with matter and generalized time
varying dark energy. Then we can write the Friedmann equation as
\begin{equation}
  H^2(z)
  =
  H_0^2 [\Omega_m (1+z)^3+\Omega_{\rm DE}\zeta(z,w)],
  \label{eq:wcdm-1}
\end{equation}
where the function $\zeta$ is given by
\begin{equation}
  \zeta(z,w)
  =
  \exp
  \left[
    3 \int_0^z \! \dd z' \frac{1+w(z')}{1+z'}
  \right],
  \label{eq:wcdm-2}
\end{equation}
where $w$ is in general a function of redshift but it is parametrized as
$w={\rm const.}$ or $w(z)=w_0 + w_a\frac{z}{1+z}$ where $w_0$ and
$w_a$ 
%is constant 
are constants
\cite[e.g.][]{Linder:03}.  The explicit cosmological
dependence of the ISW effect originates from $\partial_\tau
(D/a)={\mathcal H}D(f-1)$ and the matter power spectrum.
When we combine other cosmological data sets such as galaxy 
clustering or weak lensing, the ISW effect 
%carries an unique information of dark energy by 
provides information on dark energy via
${\mathcal H}D(f-1)$ term.
%Figure \ref{fig:isw_cosmology} shows the dependence of the ISW effect 
%on cosmological models as a function of redshift.
This quantity is shown in Figure \ref{fig:isw_cosmology} as a function
of redshift.
Less amount of dark energy makes Universe more close to the EdS
Universe, and thus produces weaker ISW effect (the black solid and the
red dashed lines).  Larger value of equation of state,  i.e. $w>-1$
(the green dot-dashed line), makes dark energy prominent at earlier
time which enhances the ISW effect.
%To put constraints on such 
Such an early dark
energy models can be tested by the cross correlation with the high-z
density tracers like quasars, which lie at $z<3$ \cite{Rossetal:12}.

Table \ref{table:comparison} shows the summary of 
%detection 
reported detections
of the
ISW effect and the constraints on dark energy models.  Boughn and
Crittenden 2002 \cite{BoughnCrittenden:02} put the first constraint on dark
energy parameter by cross-correlating the NVSS galaxies number counts
with the CMB temperature observed by the COBE.  Although they do not find
a significant detection of the ISW effect, it gives 
an
upper limit of
the amount of dark energy, $\Omega_\Lambda < 0.74$.  The redshift
distribution of the NVSS galaxies is inferred from an integral of a
luminosity function 
%for 
of
the radio galaxies \cite{DunlopPeacock:90}.
With the inferred redshift distribution, the galaxy auto-correlation
power spectrum can nicely account for the observed clustering of the
radio galaxies but it is not sufficient to validate the true
redshift distribution.  Fosalba \& Gaztanaga 2004 
\cite{FosalbaGaztanaga:04} present more stringent constraints 
%for 
on
the
$\Lambda$CDM model with the APM galaxy survey and the WMAP data. 
They fixed the Hubble parameter, $\sigma_8$ and the constant bias
parameter with assuming a flat geometry and obtained 
$0.53<\Omega_\Lambda<0.86$ $(2\sigma)$. Fosalba et
al. 2003 \cite{Fosalba+:03}
%give 
find
similar result 
%for 
from 
the combined analysis
of the APM galaxies, the SDSS main galaxies and the LRG sample, 
$0.69<\Omega_\Lambda<0.87$ $(2\sigma)$.  They utilize the
cross-correlation function for the measurement of the ISW effect.
The covariance matrix is estimated in two ways: a Jack Knife
resampling and a Monte Carlo simulation.  The Jack Knife (JK) error is
consistent with the Monte Carlo (MC) on large scales while the JK
error is significantly under-estimated on scales smaller than
$\theta<2$ degrees.  Nolta et al. 2004 \cite{Nolta+:04} revisit the NVSS
cross correlation with the WMAP.  Thanks to the greater sensitivity of
the WMAP, they 
%found 
find the 2.6$\sigma$ detection of the ISW effect with
CCF, and obtain the constraint of $\Omega_\Lambda>0$ at
$2\sigma$. They also reject the closed Universe at $3\sigma$
significance in which the ISW signal shows 
a
negative cross correlation
with the large-scale structure.  The covariance matrix is estimated by
the MC simulations for randomly generated CMB spectra,
%subtracted observed amplitude of the ISW signal, 
while keeping large-scale structure
unchanged.  This may underestimate the error since the large-scale
structure and primary CMB as well as the cross correlation between
them all contribute to the covariance matrix as shown in equation
\eqref{eq:wcdm-5}.

%For the APS, a joint analysis of the ISW combined with the galaxy 
%clustering, likelihood function is constructed with,
For a joint analysis of the ISW and the galaxy clustering APS, the
logarithm of the likelihood function, $\chi^2 \equiv -2\ln {\mathcal L}$, is given by
\begin{equation}
  \chi^2
  = ( x_{\rm obs}^{i}-\langle x^i \rangle)
  [{\bf C}^{-1}]_{ij}
  (x^j_{\rm obs}-\langle x^j \rangle),
  \label{eq:wcdm-3}
\end{equation}
where $x_{\rm obs}$ is the observed APS, $C_l^{\rm gT}$, and the index
$i$ 
%represent
represents
both the angular $l$-bin and sample used, when we
use multiple samples for the large-scale structure tracers.
%\begin{equation}
%  [{\bf C}]
%  =
%  (C_l^{{\rm gT}}- \langle C_l^{{\rm gT}} \rangle)
%  (C_l^{{\rm gT}}- \langle C_l^{{\rm gT}} \rangle)
%  \label{eq:wcdm-4}
%\end{equation}
The covariance matrix ${\bf C}$ can be estimated in different ways.
The Fisher matrix formalism gives the theoretical estimate of
the covariance matrix, while the Jack Knife (JK) or the bootstrap (BS)
resampling give the estimate of the covariance based on the
%observation 
observational
data itself.
The latter tends to underestimate the true error due to the lack of
the modes larger than the survey volume.
The
Monte Carlo simulation is particularly useful when the
%data is
data are
not Gaussian, or the 
%complicated data processes are taken such as
%complicated radial and angular selection functions.
processes such as complicated radial or angular selection functions are
taken into account.
For the Gaussian case, the covariance matrix can be simply calculated
by the Fisher formula,
\begin{equation}
  {\bf C}
  =
  \frac{1}{f_{\rm sky}(2l+1)}
  [
  \tilde{C}_l^{\rm g} \tilde{C}_l^{\rm T} + (C_l^{\rm gT})^2
  ],
  \label{eq:wcdm-5}
\end{equation}
where $\tilde{C}^{\rm g}$ is the power spectrum of the 
%galaxies distribution with the noise, $\tilde{C}_l^{\rm g}=C_l^{\rm
%  g}+\bar{n}_{\rm g}^{-1}$ and $\tilde{C}_l^{\rm T}=C_l^{\rm
%  T}+N_l^{\rm T}$ where $\bar{n}_{\rm g}$ is the number density of
%galaxies, $N_l^{\rm T}$ is the noise spectrum of CMB
%\cite[e.g.][]{Knox:95}.  
galaxies with the shot noise, 
$\tilde{C}_l^{\rm g}=C_l^{\rm g}+\bar{n}_{\rm g}^{-1}$
where $n_g$ is the number density of galaxies and 
$\tilde{C}_l^{\rm T}=C_l^{\rm T}+N_l^{\rm T}$ where
$N_l^{\rm T}$ is the noise power spectrum of the CMB.
Padmanabhan et al. 2005 \cite{Padmanabhan+:05}
use the fourth data release of the SDSS LRG photometric sample with
accurate photometric redshifts $\Delta z_p = 0.03$.  The
cross-correlation APS is used as an estimator and they introduce a
quadratic estimator \cite{Tegmark:97, Seljak:98} which enables us a
nearly maximum likelihood estimation.  Gaztanaga et
al. 2006 \cite{Gaztanaga+:06} explore wCDM model with multiple large-scale
structure tracers: 2MASS, APM, SDSS, NVSS and HEAO. They point out
that the parameter degeneracy between $\Omega_\Lambda$ and $\Omega_m$
obtained from the ISW effect is perpendicular to those from SNIa
observation and thus the combination of the ISW effect with other
cosmological probes can be a powerful tool to 
%constraint on those parameters. 
constrain those parameters.
%They obtain the cosmological constraints of
%$\Omega_\Lambda=0.7\pm 0.05$ and $w=-1.02\pm 0.17$ combined with the
%SNIa data. 
Combining the ISW data with the SNIa data, they obtain the
cosmological constraints of $\Omega_\Lambda=0.7\pm 0.05$ and 
$w=-1.02\pm 0.17$.
Using the same data set, Corasaniti et al. 2005
\cite{Corasaniti+:05} extend the analysis to constraining the model
where dark energy has clustering with a finite sound speed but 
%no constraint on the sound speed of dark energy is obtained. 
do not find a meaningful constraint on the sound speed of dark energy.
Some
works \cite{Pietrobon+:06,Vielva+:06,McEwen+:07} use wavelet for
constraining dark energy models. As we mentioned before, the wavelet
analysis has the advantage of detecting localized signals. Therefore, it has
the higher significance than other method in terms of the signal to
noise ratio of finding a signal at a given scale \cite{Vielva+:06}.
They find the signal of the ISW effect with higher significance of
$\sim3\sigma$ than other methods. The error covariance for the wavelet
is derived in a straightforward manner using equations
\eqref{eq:xcorr-16} and \eqref{eq:wcdm-5},
\begin{equation}
  \Delta 
  \left[ C_{\Psi}^{\rm ISW-X, TH}(R) \right]^2
  =
  \sum_{l}\frac{2l+1}{16\pi^2}p_l^4\Psi_l^4(R)
    \left( b_l^{\rm CMB} b_l^{\rm X} \right)^2
  \left[ C_l^{\rm CMB}C_l^{\rm X}+\left( C_l^{\rm ISW-X}\right)^2 \right].
\end{equation}
They present consistent results with the current $\Omega_\Lambda$ 
constraint \cite[e.g.][]{Planck16:13} but slightly lower values are 
favored. The equation of state parameter favors larger (less 
negative) value but it is still consistent with a cosmological
constant within $1\sigma$ level.

%%%-----------------------------------------------------------------%%%
\subsection{Non Gaussianity}
\label{sSec:ng}
%%%-----------------------------------------------------------------%%%
\begin{figure}
  \begin{center}
    \includegraphics[width=0.5\linewidth]{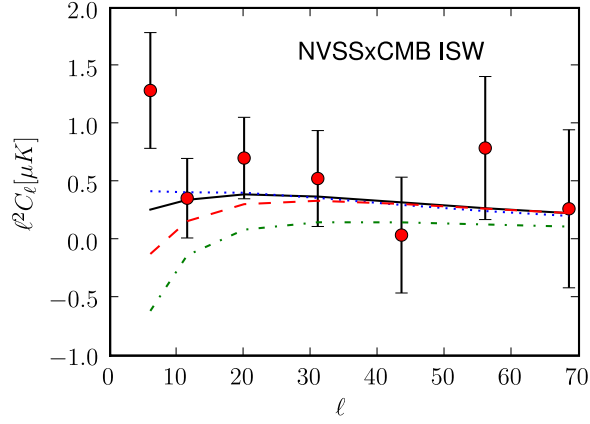}
    \caption{
      The cross correlation power spectrum of the ISW and NVSS galaxy.
      The lines show the best fit $\fnl$ model (black solid), 
      $\fnl=100$ (blue dotted), $\fnl=-100$ (red dashed), and 
      $\fnl=800$ (green dot-dashed) while keeping 
      the
      other cosmological
      parameters unchanged.
      The figure is adopted from Slosar et al. 2008 \cite{Slosar+:08}.
      \label{fig:NG}
    }
  \end{center}
\end{figure}
Although the recent observation of the CMB by the Planck 
%claims 
shows
that the non 
Gaussianity (NG) of the primordial fluctuation is consistent with zero, it is 
still worth discussing it for the future observations which 
%make more precise and accurate measurements.
would give much tighter constraints on (or potentially detect) NG.
In this section, we describe the possible impacts of the NG
of the primordial fluctuation on the ISW effect.
We assume that the NG has the form of \cite{Gangui+:94,KomatsuSpergel:01}, 
\begin{equation}
  \Phi
  =
  \phi + \fnl (\phi^2 - \langle \phi^2 \rangle)
\end{equation}
with the parameter $\fnl$ which describes the amplitude
of NG of the primordial fluctuation, 
%where $\phi$ is the random Gaussian variable. 
and $\phi$ is a random Gaussian variable.

The ISW effect basically plays two roles in the context of NG.
First, the ISW effect can be a proxy of mass of the large-scale
structure especially on large scales.
It is difficult to discriminate the primordial NG from those 
which
originate 
from the non linear gravitational evolution of the structure. However, 
the non-zero $\fnl$ alters the number of rare objects formed
responding to the initial gravitational potentials.  The number of
rare objects is enhanced if the $\fnl$ is positive but depressed for
the negative $\fnl$. Thus the effect of non-zero $\fnl$ on the halo or
galaxy power spectrum becomes prominent on large scales \cite{Dalal+:08}.
As a result, the NG bias acquires a correction of 
$b \rightarrow b + \Delta b$, where
\begin{equation}
  \Delta b(k)
  =
  2(1-b)\fnl \delta_c \frac{3\Omega_mH_0^2}{2 D(a) k^2},
\end{equation}
where $\delta_c$ is the critical density, $b$ is the usual Eulerian
constant bias.  Thus the NG can be constrained through the ISW-galaxy
cross correlation. The galaxy-galaxy auto-correlation is mainly used
for constraining the NG and 
the
ISW-galaxy cross correlation has 
%weaker ability for the constraining in terms of the statistical
%significance.
relatively lower statistical power.
However, the cross correlation is less sensitive to the systematic
effects and thus gives a robust measurement; they are complementary
to
each other.  Figure \ref{fig:NG} 
%presents 
compares
the cross correlation power
spectrum of the ISW and the NVSS galaxies with 
the 
standard $\Lambda$CDM
prediction and NG predictions keeping all 
the
parameters other than $\fnl$
unchanged \cite{Slosar+:08}. Slosar et al. 2008 \cite{Slosar+:08}
%claimed no significant detection 
found no significant detection
of the NG.
%but slightly positive $\fnl$ was favored.  
They
obtained a constraint of $\fnl = 105^{+647}_{-337}$ 
%for 
from
using only
the
ISW-NVSS cross correlation, while Afshordi \& Tolley 2008 \cite{AfshordiTolley:08}
found a 2 $\sigma$ hint for the NG, $\fnl=236\pm 127$.  Xia et
al. 2011 \cite{Xia+:11} also 
%presented 
found
a weak evidence for the
positive $\fnl$, $\fnl=74\pm 40 (1\sigma)$ 
%combined with auto
by combining the auto
correlation of the NVSS and the NVSS-ISW cross correlation.  
They used 
the
redshift distribution of the NVSS galaxies which was
directly measured by the limited number of spectroscopic galaxies
\cite{Brookes+:08} whereas Slosar et al. 2008 \cite{Slosar+:08} used one
measured from the NVSS clustering itself, which made significant 
difference in the $\fnl$.
Lately Giannantonio et al. 2014 \cite{Giannantonio+:14} point out
that the NVSS catalog contains a serious systematic error: the 
number density of galaxies depends on declination and right
ascension. This systematics largely affects auto correlation of the
NVSS galaxies while the impact of the systematics on the cross 
correlation is small.
Giannantonio and Percival 2014 \cite{GiannantonioPercival:14} revisit the NG constraints via cross 
correlation of the ISW effect with a suite of 
%galaxies samples 
galaxy samples
and
find no evidence of the NG, $\fnl=46\pm 68$ by using the ISW-galaxy 
cross correlation only.  They also take into account the cross correlation 
of the CMB-lensing with 
%galaxies samples 
galaxy samples
and 
%obtained 
obtain
a joint 
constraint of $\fnl=12\pm 21$ in the end. This is consistent with 
the CMB bispectrum analysis of the Planck \cite{PlanckXXIV:13}.

Second, the ISW effect may bring a systematic effect to the
measurement of the CMB bispectrum through the ISW-lensing correlation
\cite{GoldbergSpergel:99,PlanckXXIV:13}. The observed CMB temperature
can be decomposed into three parts; primordial CMB at last scattering
surface, lensed-CMB, and the secondary anisotropy that is generated at low
redshifts such as the ISW or the SZ. For the Gaussian field, all the
information is encoded into the power spectrum or two point
correlation function so that the bispectrum or three point function
%are 
is
the lowest statistics that describes the NG clustering.  However,
the ISW effect and the CMB-lensing are correlated 
with
each other
because the large-scale structure that makes CMB photon deflect via
gravitational lens also induces the ISW effect at low redshift.  Therefore it
mimics the primordial NG to produce a non-zero signal in the CMB
bispectrum. The ISW effect is important on large scales and
the CMB-lensing appears on much smaller scales, so the ISW-lensing
bispectrum peaks at a squeezed configuration of the triangle.  Thus
the correlation between the CMB-lens and the ISW effect brings a bias on the
measurement of the primordial NG especially for the local type by
$\Delta \fnl \simeq 10$ \cite{Komatsu:10, Cai+:10} which may
%critical for the recent accurate measurements
%be subtracted from the measurements
seriously affect the current measurement; i.e. $\sigma_{\fnl}=5.8$
\cite{PlanckXXIV:13}.  On the other hand, the ISW-lensing bias to the
other configurations of the bispectrum is negligible 
\cite{SerraCooray:08,Hanson+:09, SmithZaldarriaga:11,JunkKomatsu:12,MangilliVerde:09,Mangilli+:13}.
\section{Summary}
\label{Sec:summary}
%%%-----------------------------------------------------------------%%%
%%%-----------------------------------------------------------------%%%
In this article we review the ISW effect induced by linear and
non-linear structures of the Universe. It is well known
that the Sachs-Wolfe effect and the ISW effect are
simultaneously derived from the cosmological perturbation theory.  
%We see that the temperature anisotropy is potentially generated 
The ISW effect can be generated 
by scalar, vector and tensor mode fluctuations of the metric but the
significant contribution comes from 
the
scalar mode, which is related to
the time variation of the matter density fluctuation. 
%However, 
%the recent observation of BICEP2 suggests the large amplitude of
%the tensor mode, i.e. $r=0.1-0.2$. The tensor mode with large $r$
%value is still subdominant to the scalar mode but it enhances the
%CMB temperature fluctuation on large angular scales and thus 
%affect the cosmological model constraints, e.g. the negative running 
%of scalar spectral index, or the existence of the isocurvature
%fluctuations are preferred.
The recent detection of the tensor-mode CMB polarization claimed by
the BICEP2 collaboration \cite{BICEP2:14} can indeed be the
polarization generated by the tensor-mode ISW effect from primordial
gravitational waves.
%For the scalar mode perturbation, there are interpretations of the
%time variation 
The time variation
of the gravitational potential in the standard 
$\Lambda$CDM Universe
%,
comes from:
1) the coherent decay of the
gravitational potential due to the accelerating expansion of the
Universe, 2) the isotropic clustering inflow of mass lump toward
the center of 
%the mass 3) 
mass, and 3)
the transverse motions of the clusters. The
latter two effects are 
%so tiny that we can not detect 
too tiny to detect
with the current
CMB experiments but it might be possible to detect with
future experiments having finer angular resolutions such as
ACTPol \cite{ACTPol:10}, SPTPol \cite{SPTPol:12} or COrE \cite{COrE:11}.

%The observational studies for the ISW effect has been also explored. 
We also review the observational studies of the ISW effect.
%It is 
The ISW effect was
first detected by the cross correlation of the CMB
observed by the WMAP with the number counts of the radio galaxies
measured by the NVSS. The significance of detection 
%is 
was
$2-3\sigma$
\cite{BoughnCrittenden:04}.  Subsequently a number of 
%works have been done 
detections have been reported
with various tracers of dark matter 
%in 
using a
variety of
statistical methods. As the ISW effect reflects the large-scale
fluctuations in the Universe, they can be used for constraining the
cosmological models especially at low redshifts ($z<1$).  The
statistical error of the ISW effect is mostly dominated by sample
variance but the detection significances vary among papers and are often
not consistent with each other. It is partly due to the consequence of the
different statistics they used, 
%the different treatment 
different treatments 
of the
foreground, or different estimates or assumptions on
the redshift distribution of galaxies. 
%As we have progressively acquired the knowledge for fixing those
%issues, and enough large volume of data, we need more progresses 
%to wholly understand the
As we acquire more knowledge on those issues, we may be able to
understand 
inconsistencies seen in table \ref{table:comparison}.  
All sky polarization data from the Planck will be useful for further 
studying the dust model of our Galaxy  as well as the confirmation
of the large amplitude of tensor mode fluctuation discovered by the
BICEP2.
In addition to this, complete BOSS spectroscopic galaxies and QSOs
samples can be used for calibrating the redshift distribution of
galaxies and QSOs.
Therefore both the Planck and the BOSS data which are going to be
delivered soon 
would
allow us to extensively study the ISW effect with
better  understandings of the systematics.

%%%-----------------------------------------------------------------%%%
%%%-----------------------------------------------------------------%%%
\section*{acknowledgement}
The author would like to thank Eiichiro Komatsu for the careful
reading of this article and giving 
%substantial 
many useful comments.
This work is supported in part by the FIRST program "Subaru Measurements 
of Images and Redshifts (SuMRe)" initiated by the Council for Science 
and Technology Policy (CSTP).
%%%-----------------------------------------------------------------%%%
%%%-----------------------------------------------------------------%%%

%%%-----------------------------------------------------------------%%%
%%%-----------------------------------------------------------------%%%
%\bibliographystyle{ptephy}
%\bibliography{ms}
%%%-----------------------------------------------------------------%%%
%%%-----------------------------------------------------------------%%%

\end{document}